\documentclass[preprint,authoryear,review,10pt]{elsarticle}

\usepackage{natbib}
\usepackage{color}
\usepackage{times}
\usepackage{latexsym}
\usepackage{enumerate}
\usepackage{epsfig}
\usepackage{amsmath}
\usepackage{amssymb}
\usepackage{amsfonts}
\usepackage[english]{babel}
\usepackage{graphicx}
\usepackage{graphics}
\usepackage{psfrag}
\usepackage{amsbsy}
\usepackage{amsmath}
\usepackage{textcomp}
\usepackage{url}
\usepackage[latin1]{inputenc} 
\usepackage{array} 
\usepackage{arydshln} 
\usepackage{multirow} 
\usepackage{subfigure} 
\usepackage{cancel} 
\usepackage{ctable}
\usepackage{slashbox}
\newcommand{\real}{\text{I\!R}}

\def\bx{\boldsymbol{x}}
\def\by{\boldsymbol{y}}
\def\bz{\boldsymbol{z}}

\def\bw{\boldsymbol{w}}

\def\bX{\boldsymbol{X}}
\def\bY{\boldsymbol{Y}}
\def\bW{\boldsymbol{W}}

\def\b0{\boldsymbol{0}}
\def\b1{\boldsymbol{1}}

\def\cL{\mathcal{L}}

\def\mN{\mathbb{N}}
\def\EE{\mathbb{E}}

\def\balpha{\boldsymbol{\alpha}}

\def\bmu{\mbox{\boldmath $\mu$}}
\def\bkappa{\mbox{\boldmath $\kappa$}}

\def\bpsi{\mbox{\boldmath $\psi$}}

\def\bbeta{\mbox{\boldmath $\beta$}}

\def\bxi{\mbox{\boldmath $\xi$}}
\def\btheta{\mbox{\boldmath $\theta$}}

\def\bphi{\mbox{\boldmath $\phi$}}
\def\bvarphi{\mbox{\boldmath $\varphi$}}
\def\bpi{\mbox{\boldmath $\pi$}}

\def\bzeta{\mbox{\boldmath $\zeta$}}

\def\bSigma{\mbox{\boldmath $\Sigma$}}

\def\sbX{\underset{\widetilde{}}{\bX}}

\newtheorem{Prop}{Proposition}

\newtheorem{Cor}[Prop]{Corollary}

\newtheorem{Ex}{Example}
\bibpunct{(}{)}{,}{a}{}{;}

\journal{}

\begin{document}
\begin{frontmatter}



\title{Generalized Linear Gaussian Cluster-Weighted Modeling}

\author[Ingrassia_Punzo]{Salvatore Ingrassia}
\author[Minotti_Vittadini]{Simona C. Minotti}
\author[Ingrassia_Punzo]{Antonio Punzo}
\author[Minotti_Vittadini]{Giorgio Vittadini}
\address[Ingrassia]{Dipartimento di Economia e Impresa, Universit\`{a} di Catania (Italy) \\
    Corso Italia, 55 - 95129 Catania (Italy), [s.ingrassia, antonio.punzo]@unict.it}
\address[Minotti_Vittadini]{Dipartimento di Statistica, Universit\`{a} di Milano-Bicocca (Italy) \\
		Via Bicocca degli Arcimboldi, 8 - 20126 Milano (Italy)
              [simona.minotti, giorgio.vittadini]@unimib.it}


\begin{abstract} 
Cluster-Weighted Modeling (CWM) is a flexible mixture approach for modeling the joint probability of data coming from a heterogeneous population as a weighted sum of the products of marginal distributions and conditional distributions. 
In this paper, we introduce a wide family of Cluster Weighted models in which the conditional distributions  are assumed to belong to the exponential family with canonical links which will be referred to as \textit{Generalized Linear Gaussian Cluster Weighted Models}. 
Moreover, we show that, in a suitable sense, mixtures of generalized linear models can be considered as nested in Generalized Linear Gaussian Cluster Weighted Models. 
The proposal is illustrated through many numerical studies based on both simulated and real data sets.
\end{abstract}

\begin{keyword}
Cluster-Weighted Modeling, Mixture Models, Model-Based Clustering, Generalized Linear models.


\end{keyword}

\end{frontmatter}

\section{Introduction}
\label{sec:Introduction}

Let $\left(\bX',Y\right)'$ be a random vector, defined on $\Omega$, composed by a $d$-dimensional explanatory variable $\bX$ and a unidimensional response variable $Y$.
Let $p\left(\bx,y\right)$ be the joint distribution of $\left(\bX',Y\right)'$.
Suppose that $\Omega$ can be partitioned into $G$ groups, say $\Omega_1,\ldots,\Omega_G$.
{\em Cluster-Weighted Modeling} (CWM) represents a convenient mixture approach for modeling data of this type.
Indeed, in this approach, the joint probability of $\left(\bX',Y\right)'$ can be written as
\begin{equation}
p(\bx,y;\btheta)=
\sum^{G}_{g=1} p(y|\bx,\Omega_g) p(\bx|\Omega_g) p(\Omega_g) =
\sum^{G}_{g=1} p(y|\bx;\bxi_g) p(\bx;\bpsi_g) \pi_g, 
\label{eq:CWM base}
\end{equation}
where $p(y|\bx,\Omega_g)=p(y|\bx;\bxi_g)$ is the conditional distribution of $Y$ given $\bx$ in $\Omega_g$ (depending on some parameters $\bxi_g$), $p(\bx|\Omega_g)=p(\bx;\bpsi_g)$ is the probability distribution of $\bx$ in $\Omega_g$ (depending on some parameters $\bpsi_g$), $\pi_g=p(\Omega_g)$ is the mixing weight of $\Omega_g$ ($\pi_g > 0$ and $\sum_g \pi_g =1$), and $\btheta=\left\{\bxi_g,\bpsi_g,\pi_g;g=1, \ldots, G\right\}$ denotes the set of all model parameters.

In the original formulation of {\em Cluster Weighted Modeling}(CWM), the random vector $\left(\bX',Y\right)'$ is assumed to be real-valued, see \citet{Gers:Nonl:1997} and also \citet{Gers:Nature:1999},  \citet{Schoner:2000} for details).
Quite recently, \citet{Ingr:Mino:Vitt:Loca:2012} reformulated CWM  in a statistical setting showing that it is a quite general and flexible family of mixture models. 
Also, in the majority of the existence literature about CWM, a linear relationship of $Y$ on $\bx$ is considered for each $\Omega_g$, that is
\begin{equation}
Y=\mathbb{E}(Y|\bx,\Omega_g)+\varepsilon_g=\mu(\bx;\bbeta_g)+\varepsilon_g=\beta_{0g}+\bbeta'_{1g}\bx+\varepsilon_g,
\label{eq:linear relationship}
\end{equation}
where $\bbeta_g=(\beta_{0g},\bbeta'_{1g})'$, with $\bbeta_g\in \real^{d+1}$, and $\varepsilon_g$ is assumed to have zero mean and a finite variance.
If in each $\Omega_g$, a Gaussian distribution is assumed for both $Y|\bx$ and $\bX$, say $Y|\bx\sim N(0,\sigma^2_g)$ and $\bX \sim N_d(\bmu_g,\bSigma_g)$, then the original linear Gaussian CW model is obtained in \citet{Gers:Nonl:1997}.
For robustness sake, \citet{Ingr:Mino:Vitt:Loca:2012} also introduce linear $t$ CWM, which is based on the assumption of a (multivariate) $t$ distribution for both $Y|\bx$ and $\bX$ in each mixture component.
Starting from this result, \citet{Ingr:Mino:Punz:Mode:2012} define a family of twelve linear $t$ CWMs for model-based clustering and classification.
Finally, by generalizing \eqref{eq:linear relationship} via a polynomial relationship,  \citet{Punz:Flex:2012} defines the polynomial Gaussian CWM.   

However, in many cases we have to face with modeling categorical variables depending on numerical covariates based on data coming from a heterogeneous population. 
For example, in effectiveness studies based on administrative data, one may be interested to estimate many different effectiveness models for certain groups of users, taking into account the characteristics of each group; in the healthcare context, one typical outcomes are mortality rate or  length of stay, while in the education framework an outcome may be the success in an exam.  
In statistical literature, such problems are usually approached considering {\em finite mixtures of generalized linear models}
(FMGLM), see e.g. \citet{McLa:1997}, \cite{McLa:Peel:fini:2000}, \citet{Wede:DeSa:mixtglm:1995}. 

This paper generalizes the linear Gaussian CWM by considering a generalized linear model (with canonical link) for the relationship of $Y$ on $\bx$ in each $\Omega_g$, $g=1,\ldots,G$. In particular, 
we introduce a wide family of Cluster Weighted (CW) models in which the conditional distributions  are assumed to belong  to the exponential family with canonical links which will be referred to as {\em Generalized Linear Gaussian Cluster Weighted Models (GLGCWM)}. 
We remark that while FMGLM model conditional distributions, Cluster Weighted approaches model the joint distribution as the product of marginal  and conditional distributions. 
In this paper, we show also that, in some sense, FMGLM can be considered as nested models of GLGCWM.

Generalized linear models (with canonical link) are regression models where $Y$ is specified to be distributed according to one of the members of the exponential family. 
Accordingly, those models deal with dependent variables $Y$ that can be either continuous with, for example, Gaussian, gamma or inverse Gaussian distributions, or discrete with, for example, binomial or Poisson distributions.
The exponential family is a very useful class of distributions and the common properties of the distributions in the class enable them to be studied simultaneously rather than as a collection of unrelated cases. 
GLGCWM thus affords a general framework that, in addition to encompass the linear Gaussian CWM as a special case, allow us to use the CW principle also for discrete dependent variables, which are very common in real applications (see, e.g., \citealp{Wed:DeSar:bin:1993, Wed:DeSar:Bul:Ram:poi:1993, Wan:Put::mix:1998, WCP:anal:1998, Wan:Yau:Lee:poi:2002, Yan:Lai:poi:2005, Xia:Yau:Lee:Fun:infl:2005}).

We remark that also \citet{Gers:Nature:1999} coped with the problem of discrete set of values such as events, patterns or conditions, but they did not really model the joint probability of the dependent variable and the covariates; what they introduced in the CWM is the histogram of the probabilities of each state in the clusters (without explicit dependence on the covariates).

The remainder of the paper is organized as follows. 
In Section~\ref{sec:The Generalized Linear Gaussian Cluster-Weighted Model}, we introduce the Generalized Linear Gaussian CWM (GLGCWM);
in Section~\ref{sec:relationships}, we show some results about the relationships of the GLGCWM with mixture of generalized linear models (with and without concomitants);
in Section~\ref{sec:Parameter estimation}, we present parameter estimation according to likelihood approach;
in Section~\ref{sec:EM}, we describe the EM algorithm for parameter estimation in GLCWM;
in Section \ref{sec:Performance evaluation}, we introduce some  measures for performance evaluation;
in Section~\ref{sec:numerical studies}, we illustrate numerical studies based on both simulated and real data. 
Finally, conclusions and perspectives for future research are presented in Section~\ref{sec:Conclusions}.
\section{The model}
\label{sec:The Generalized Linear Gaussian Cluster-Weighted Model}

We consider a broad family of models in which, for each $\Omega_g$, the conditional distributions are assumed to belong to the exponential family with canonical links, that is 
\begin{equation}
p(y|\bx;\bxi_g )= q(y|\bx; \bbeta_g, \lambda_g) = \exp\left\{\frac{y \eta(\bx; \bbeta_g) - b\left[\eta(\bx; \bbeta_g)\right]}{a(\lambda_g)} + c(y,\lambda_g)\right\}, 
\label{eq:canonical exponential family}
\end{equation}
where $a(\cdot)$, $b(\cdot)$, and $c(\cdot)$ are specified functions, $\lambda_g$ is the \textit{dispersion parameter}, with $a(\lambda_g)>0$, and $\eta(\bx; \bbeta_g)=\eta_g=\beta_{0g}+\bbeta'_{1g}\bx$ is the {\textit{canonical function}} (see, e.g., \citealp{Wede:DeSa:mixtglm:1995} and \citealp{McCullaghNelder:GLM:2000}). 
In particular, $b(\cdot)$ is the \textit{cumulant function}, while $a(\cdot)$ and $b(\cdot)$ satisfy
\begin{displaymath}
\mu_g = \EE(Y|\bx;\bbeta_g,\lambda_g) = b'(\eta_g)
\quad \text{and} \quad 
\sigma^2_g  = \mathbb{V}(Y|\bx;\bbeta_g,\lambda_g) = a(\lambda_g) b''(\eta_g), 
\end{displaymath}
where $b'(\eta_g)$ and $b''(\eta_g)$ are the first and second derivatives of $b(\eta_g)$ with respect to $\eta_g$, respectively.
Moreover, $\eta_g$ is related to the expected value $\mu(\bx;\bbeta_g)=\mu_g$, through a link function $h(\cdot)$, in the following way
\begin{equation*}
\eta_g = h(\mu_g).
\end{equation*}
For sake of simplicity, we assume Gaussian distributions for marginals, i.e. $\bX|\Omega_g \sim N_d(\bmu_g,\bSigma_g)$.
Thus, equation \eqref{eq:CWM base} becomes
\begin{equation}
p(\bx,y;\btheta)=\sum^{G}_{g=1} q(y|\bx;\bbeta_g, \lambda_g)\phi_d(\bx;\bmu_g,\bSigma_g) \pi_g .
\label{eq:Generalized Linear Gaussian CWM}
\end{equation}
Model \eqref{eq:Generalized Linear Gaussian CWM} will be referred to as \textit{generalized linear Gaussian CWM} (GLGCWM) hereafter.
Canonical links are the identity, log, logit, inverse and squared inverse functions for the normal, Poisson, binomial, gamma and inverse Gaussian distributions \citep[see][\tablename~2.1]{McCullaghNelder:GLM:2000}.
Thus, all these distributions can be taken into account for modeling the $Y$ variable in each $\Omega_g$.
In particular, the Poisson and the Binomial distributions are quite useful because they allow the CW principle to be applied also for discrete response variables $Y$; this is the reason why, in the following, we shall focus mainly on such distributions.

It is interesting to note that, from a clustering/classification point of view, the posterior probabilities of group membership
\begin{equation}
p(\Omega_g|\bx,y;\btheta) =  
\frac{p(y|\bx;\bbeta_g, \lambda_g) p(\bx;\bmu_g, \bSigma_g) \pi_g}{p(\bx,y;\btheta)}, \quad g=1,\ldots,G,
\label{eq:CWM posteriors}
\end{equation}
depend on both the marginal and conditional component distributions, differently from the standard mixture models.


\subsection{The Binomial Gaussian CWM}
\label{subsec:The Binomial Gaussian CWM}

Assume that $Y$ takes values in $\{0,1, \ldots, M\}$, for some  $M \in \mN$.
Moreover, assume that the probability mass function of $Y|\bx$ in $\Omega_g$ is  binomial with parameters $(M, \gamma_g(\bx))$, that is $Y|\bx,\Omega_g \sim \text{Bin}(M,\gamma_g(\bx))$.
In this case
\begin{equation}
q(y|\bx;M,\beta_{0g},\bbeta_g) = \binom{M}{y} [\gamma_g(\bx)]^y\left[1-\gamma_g (\bx )\right]^{M-y},  
\label{eq:Binomial Y}
\end{equation}
where the probability $\gamma_g(\bx )$ is postulated to depend on $\bx$ through the function 
\begin{equation} 
\gamma_g (\bx )= \frac{\exp(\beta_{0g} + \bbeta_g'\bx)}{1+\exp(\beta_{0g} + \bbeta_g'\bx)} \quad \text{or, equivalently,} \quad \ln\left(\frac{\gamma_g (\bx )}{1-\gamma_g  (\bx )}\right) = \beta_{0g} + \bbeta_g'\bx, 
\label{eq:Binomial link}
\end{equation}
with $\beta_{0g} \in \real$ and  $\bbeta_g \in \real^d$ being parameters to be estimated. 
When \eqref{eq:Binomial Y} is considered in \eqref{eq:CWM base}, we have \textit{Binomial Gaussian CWM} or, more simply, \textit{Binomial CWM}.
Then, \textit{Bernoulli CWM} results, as a special case, when $M=1$. 


\subsection{The Poisson Gaussian  CWM}
\label{subsec:The Poisson Gaussian CWM}

Assume that $Y$ takes values in $\mN$.
Moreover, assume that the probability mass function of $Y|\bx$ in $\Omega_g$ is Poisson with parameter $\lambda_g (\bx )$, that is $Y|\bx,\Omega_g \sim \text{Poi}(\lambda_g (\bx ))$.
In this case
\begin{equation}
q(y|\bx;\beta_{0g},\bbeta_g) = \exp\left[-\lambda_g(\bx)\right]\frac{[\lambda_g(\bx)]^y}{y!},
\label{eq:Poisson Y}
\end{equation}
where $\lambda_g(\bx)$ is postulated to depend on $\bx$ through the function 
\begin{equation} 
\lambda_g (\bx)= \exp(\beta_{0g} + \bbeta_g'\bx) \quad \text{or, equivalently,} \quad \ln[\lambda_g (\bx)]  = \beta_{0g} + \bbeta_g'\bx, 
\label{eq:Poisson link}
\end{equation}
with $\beta_{0g} \in \real$  and  $\bbeta_g \in \real^d$  being parameters to be estimated. 
When \eqref{eq:Poisson Y} is considered in \eqref{eq:CWM base}, we have \textit{Poisson Gaussian CWM} or, more simply,
\textit{Poisson CWM}.

\section{Relationships with finite mixtures of generalized linear models}
\label{sec:relationships}

In this section, we extend  results given in \citet{Ingr:Mino:Vitt:Loca:2012}, \citet{Ingr:Mino:Punz:Mode:2012}  to the generalized linear Gaussian CWM. 
Proofs are given in the Appendix.
Let
\begin{equation}
f(y|\bx;\bkappa) = \sum^{G}_{g=1}q(y|\bx;\bbeta_g, \lambda_g)\pi_g  
\label{eq:finite mixture of GLMs}
\end{equation}
be a {\em finite mixture of generalized linear model} (FMGLM), see e.g.  \citet{McLa:1997}, \citet{McLa:Peel:fini:2000}, \citet{Wede:DeSa:mixtglm:1995} where  $\bkappa=\{\bbeta_g, \lambda_g,\pi_g; g=1,\ldots,G\}$ denotes the overall parameters of the model.
According to this model, the posterior probability that the generic observation $(\bx',y)'$ belongs to $\Omega_g$ is 
\begin{equation}
p(\Omega_g|\bx,y) = \frac{q(y|\bx;\bbeta_g, \lambda_g)\pi_g}{f(y|\bx;\bkappa)},  
\quad g=1, \ldots, G. 
\label{eq:posteriors from a finite mixture of GLMs}
\end{equation}

\begin{Prop}
\label{prop:CWM vs finite mixture of GLMs}{\rm 
If the marginal distributions $\phi_d(\bx;\bmu_g, \bSigma_g)$ of $\bX|\Omega_g$ do not depend on $\Omega_g$, i.e.
$\bmu_g, \bSigma_g=\bmu, \bSigma$, $g=1,\ldots,G$, then model \eqref{eq:Generalized Linear Gaussian CWM} becomes
\begin{equation*}
p(\bx,y;\btheta) = \phi_d(\bx;\bmu, \bSigma) f(y|\bx; \bkappa),
\end{equation*}
where $f(y|\bx;\bkappa)$ was defined in \eqref{eq:finite mixture of GLMs}. 
}\end{Prop}

\begin{Cor}[from Proposition~\ref{prop:CWM vs finite mixture of GLMs}]
\label{cor:posteriors CWM vs posteriors finite mixture of GLMs}{\rm
Under the assumptions of Proposition~\ref{prop:CWM vs finite mixture of GLMs}, the posterior probability that the generic observation $(\bx',y)'$ belongs to $\Omega_g$ from model \eqref{eq:Generalized Linear Gaussian CWM} coincides with \eqref{eq:posteriors from a finite mixture of GLMs}.
}\end{Cor}

\paragraph{\bf Remark}
We remark that  result in Proposition \ref{prop:CWM vs finite mixture of GLMs} holds in a quite more general context. As a matter of fact, the proof
does not require any distributional assumption on the marginal distributions, but it needs  only that 
 the marginal distributions  $\bX|\Omega_g$ do not depend  on group $g$, i.e.  
$\bpsi_g =\bpsi$ for every $g=1,\ldots,G$. 
Thus, the model \eqref{eq:CWM base} yields:
\begin{equation*}
p(\bx,y; \btheta) = p(\bx;\bpsi)  f\left(y|\bx; \bkappa\right),
\end{equation*}
where $f(y|\bx; \bphi)$ has been defined in \eqref{eq:finite mixture of GLMs}. 

An extension of model \eqref{eq:finite mixture of GLMs} concerns the case with concomitant variable; this leads to 
{\em finite mixtures of GLMs with concomitant variables} (FMGLMC), see \citet{Grun:Leis:Flex:2008} 
\begin{equation}
f(y|\bx;\bvarphi) = \sum^{G}_{g=1} q(y|\bx; \bbeta_g, \lambda_g) p(\Omega_g|\bx;\balpha_g), \label{eq:finite mixture of GLMs with concomitant}
\end{equation}
where  the mixing weight $p(\Omega_g|\bx;\balpha_g)$ is now a function of $\bx$ through some parameters $\balpha$, and $\bvarphi$ denotes the overall parameters of the model.
Note that, in the general formulation of model \eqref{eq:finite mixture of GLMs with concomitant} given by \citet[][p.~3]{Grun:Leis:Flex:2008}, the concomitant variables could be also exogenous to $\bX$.    
The probability $p(\Omega_g|\bx;\balpha_g)$ is usually modeled by a multinomial logistic distribution with the first component as baseline, that is
\begin{equation}
p(\Omega_g|\bx;\balpha_g) = \frac{\exp(\alpha_{g0}+\balpha_{g1}'\bx)}{\displaystyle\sum^{G}_{j=1} \exp(\alpha_{j0}+\balpha_{j1}'\bx)}, 
\label{mult_log}
\end{equation}
where $\balpha_g=(\alpha_{g0},\balpha_{g1}')'$.
According to model \eqref{eq:finite mixture of GLMs with concomitant}, with the specification of $p(\Omega_g|\bx;\balpha_g)$ given in \eqref{mult_log}, the posterior probability that the generic observation $(\bx',y)'$ belongs to $\Omega_g$ is 
\begin{equation}
p(\Omega_g|\bx,y) = \frac{q(y|\bx;\bbeta_g, \lambda_g) p(\Omega_g|\bx; \balpha_g)}{\displaystyle\sum_{j=1}^G q(y|\bx;\bbeta_j, \lambda_j) p(\Omega_j|\bx;\balpha_j)} = 
\frac{q(y|\bx;\bbeta_g, \lambda_g) \exp(\alpha_{g0}+\balpha_{g1}'\bx)}{\displaystyle\sum_{j=1}^G q(y|\bx;\bbeta_j, \lambda_j) \exp(\alpha_{j0}+\balpha_{j1}'\bx)}. 
\label{eq:posterior from finite mixture of GLMs with multinomial concomitant}
\end{equation}

\begin{Prop}
\label{prop:CWMLG-FMLRC}{\rm 
If in \eqref{eq:Generalized Linear Gaussian CWM} it results $\bX|\Omega_g \sim N_d(\bmu_g,\bSigma)$
and $\pi_g=\pi=1/G$, $g=1,\ldots,G$, then
\begin{equation*}
p(\bx,y; \btheta) = p(\bx;\bpsi) f(y|\bx;\bvarphi),
\end{equation*}
where $f(y|\bx;\bvarphi)$ is defined in \eqref{eq:finite mixture of GLMs with concomitant} through \eqref{mult_log} and $p(\bx;\bpsi) = 1/G \displaystyle\sum_{g=1}^G \phi_d(\bx;\bmu_g,\bSigma)$.
} 
\end{Prop}

\begin{Cor}[from Proposition~\ref{prop:CWMLG-FMLRC}]
\label{cor:CWMLG-FMLRC}{\rm
Under the assumptions of Proposition~\ref{prop:CWMLG-FMLRC}, the posterior probability that the generic observation $(\bx',y)'$ belongs to $\Omega_g$ from model \eqref{eq:Generalized Linear Gaussian CWM} coincides with \eqref{eq:posterior from finite mixture of GLMs with multinomial concomitant}.
}\end{Cor}
\section{Likelihood function and parameter estimation}
\label{sec:Parameter estimation}

Let ${(\bx_1,y_1), \ldots, (\bx_N,y_N)}$ be a sample of $N$ independent observation pairs drawn from model in (\ref{eq:CWM base})
 and set $\sbX=(\bx'_1, \ldots, \bx'_n)$, $\bY=(y_1, \ldots, y_n)$. 
The likelihood function of the generalized linar Gaussian CWM \eqref{eq:Generalized Linear Gaussian CWM} is given by:
\begin{align*}
L_0(\btheta; \sbX, \by) & = \prod_{n=1}^N p(\bx_n,y_n;\btheta)  = \prod_{n=1}^N \left[\sum_{g=1}^G q(y_n|\bx_n; \bbeta_g, \lambda_g)
\phi_d(\bx_n;\bmu_g,\bSigma_g ) \pi_g \right]  .
 \label{likCWMg} 
\end{align*}  
Maximization of $L_0(\btheta; \bX, \by)$ with respect to $\btheta$ yields the maximum likelihood estimate of $\btheta$.
Previous results under Gaussian assumptions have been presented in \citet{Ingr:Mino:Maxi:2012}.
Let us consider fully categorized data:
\begin{equation*}
\{ \bw_n \, : \, n=1, \ldots, N \} = \{ (\bx_n, y_n, \bz_n) \, : \, n=1, \ldots, N \},
\end{equation*}
where $\bz_n = (z_{n1}, \ldots, z_{ng})'$, with  $z_{ng}=1$ if $(\bx_n,y_n)$ comes from the $g$-th population and $z_{ng}=0$ otherwise. 
Then, the complete-data likelihood function corresponding to $\bW=(\bw_1, \ldots, \bw_N)$ can be written in the form:
\begin{equation}
L_c(\btheta; \sbX, \by) = \prod_{n=1}^N \prod_{g=1}^G [q(y_n|\bx_n; \bbeta_g, \lambda_g)]^{z_{ng}} [\phi_d(\bx_n;\bmu_g, \bSigma_g ) ]^{z_{ng}}  \pi_g^{z_{ng}}. \label{likcCWMg} 
\end{equation}
Taking the logarithm of \eqref{likcCWMg} after some algebra we get:
\begin{align}
\cL_c (\btheta; \sbX, \by) & = \ln L_c(\btheta; \sbX, \by)  \notag \\
& = \sum_{n=1}^N \sum_{g=1}^G [ z_{ng} \ln q(y_n|\bx_n; \bbeta_g, \lambda_g)+ z_{ng} \ln  \phi_d(\bx_n;\bmu_g, \bSigma_g ) + z_{ng} \ln \pi_g ] \notag \\
& = \sum_{n=1}^N \sum_{g=1}^G  z_{ng} \ln q(y_n|\bx_n; \bbeta_g, \lambda_g)+ \sum_{n=1}^N \sum_{g=1}^G z_{ng} \ln  \phi_d(\bx_n;\bmu_g, \bSigma_g )  + \nonumber \\ & \qquad + \sum_{n=1}^N \sum_{g=1}^G z_{ng} \ln \pi_g \notag \\
& = \cL_{1c} (\bxi) + \cL_{2c} (\bpsi) + \cL_{3c} (\bpi), \label{llikCWM}
\end{align}
where $\bxi = \{ \bbeta_g, \lambda_g \, ; g=1, \ldots, G\}$ and $\bpsi=\{ \bmu_g, \bSigma_g \,; g=1, \ldots, G\}$.
We remark that in \eqref{llikCWM} the weights $\pi$ are estimated through the posterior probability \eqref{eq:CWM posteriors}, see Section \ref{sec:EM} for details. 

\paragraph{\bf Relationship with the log-likelihood function of FMGLM} 

In the following   we show  that, under suitable hypotheses, the maximization
of the likelihood function of GLGCWM in \eqref{eq:Generalized Linear Gaussian CWM} leads to the same parameter estimates of 
FMGLM in \eqref{eq:finite mixture of GLMs}. Indeed,  
based on \eqref{eq:finite mixture of GLMs}, the log-likelihood is given by
\begin{align}
\cL_c (\btheta; \sbX, \by)   & = \sum_{n=1}^N \sum_{g=1}^G  ( z_{ng} \ln  q(y_n|\bx_n; \bbeta_g, \lambda_g)
 + z_{ng} \ln  \pi_g ) \notag \\
 & = \sum_{n=1}^N \sum_{g=1}^G  z_{ng} \ln q(y_n|\bx_n; \bbeta_g, \lambda_g)  
 + \sum_{n=1}^N \sum_{g=1}^G z_{ng} \ln  \pi_g  \notag \\
  &= \cL_{1c} (\bxi) + \cL_{3c} (\bpi) . \label{llikcFMR}
\end{align}
\begin{Prop}\label{CWM->FMR}{\rm
In model \eqref{eq:Generalized Linear Gaussian CWM}, if the local densities $\phi_d(\bx_n;\bmu_g, \bSigma_g ) $ have the same parameters $(\bmu_g, \bSigma_g)=(\bmu, \bSigma)$  for   $g=1, \ldots, G$,
 then maximum likelihood estimate of $(\bxi, \bpi)$ in \eqref{llikcFMR} coincides with
the corresponding estimate in \eqref{llikCWM}.  
}\end{Prop}
%
\paragraph{\bf Remark} We remark that the above result can be generalized like in the previous case.
As a matter of fact, the proof of Proposition \ref{CWM->FMR} does not require the Gaussian assumption on the marginal distribution. 
For any marginal distribution $p(\cdot; \bpsi_g)$ (depending on some parameter $\psi_g$), 
if the local densities $p(\bx_n;\bpsi_g ) $ have the same parameters $\bpsi_g=\bpsi$  for   $g=1, \ldots, G$,
 then maximum likelihood estimate of $(\bxi, \bpi)$ in \eqref{llikcFMR} coincides with
the corresponding estimate in \eqref{llikCWM}.

\paragraph{\bf Relationship with the loglikelihood function of FMGLMC} 
Now let us analyze the relationships with FMGLMC. 
In particular, we show  that, under suitable hypotheses, the maximization of the likelihood function of GLGCWM in \eqref{eq:Generalized Linear Gaussian CWM} leads to the same parameter estimates of FMGLMC in \eqref{eq:finite mixture of GLMs with concomitant}. 
Indeed, let us consider the density function in \eqref{eq:finite mixture of GLMs with concomitant},
where the mixing weights $ p(\Omega_g|\bx, \balpha_g)$ are given in \eqref{mult_log}.
The corresponding complete-data log-likelihood function  is given by:
\begin{align}
\cL_c (\btheta; \sbX, \by)   & = \sum_{n=1}^N \sum_{g=1}^G  \left[ z_{ng} \ln q(y_n|\bx_n; \bbeta_g, \lambda_g) 
 + z_{ng} \ln  p(\Omega_g|\bx,\balpha_g) \right] \notag \\
  &= \cL_{1c} (\bxi) + \cL_{3c} (\bzeta)   , \label{llikcFMRC}
\end{align}
where $\bzeta = \{\balpha_g \, ; g=1, \ldots, G\}$.

\begin{Prop}{\label{CWM->FMRC}\rm
In model \eqref{eq:Generalized Linear Gaussian CWM}, assume that the local densities have  the same covariance matrices, i.e. $ \phi_d(\bx;\bmu_g, \bSigma_g )   = \phi_d(\bx; \bmu_g, \bSigma)$  and the prior probabilities be equal, i.e. $\pi_g=1/G$ for $g=1, \ldots, G$.
Then, the maximum likelihood estimate of $(\bxi, \bzeta)$ in \eqref{llikcFMRC} can be derived from  the  estimate of $(\bxi, \bpsi)$ in \eqref{llikCWM}.
}\end{Prop}

\subsection{Discussion}\label{subsec:discussionCWM->FMR}
The above results show that both models FMGLM and FMGLMC can be considered as nested models of CWM in \eqref{eq:CWM base} even if
 they have a different structure. As a matter of fact, model \eqref{eq:finite mixture of GLMs} and \eqref{eq:finite mixture of GLMs with concomitant} consider only conditional distributions, while CWM considers joint distributions (as a product of marginal and conditional distributions). However, we remark that:
\begin{itemize}
\item if the marginal distributions $p(\bx; \bmu_g, \bSigma_g)$ in \eqref{eq:CWM base} do not depend on the $g$th group, i.e. $p(\bx; \bmu_g, \bSigma_g)=p(\bx; \bmu, \bSigma)$ ($g=1, \ldots, G)$,
then the estimates $\{\bbeta_g, \lambda_g, \pi_g, \; g=1, \ldots, G\}$ in \eqref{eq:CWM base} and  \eqref{eq:finite mixture of GLMs} are the same according to Proposition \ref{CWM->FMR}; in other words, if $\bmu_g, \bSigma_g=\bmu, \bSigma$ ($g=1, \ldots, G$), then the parameters of 
the conditional distributions in GLGCWM and FMGLM  are the same. Moreover,  the posterior probability \eqref{eq:CWM posteriors} reduces to \eqref{eq:posteriors from a finite mixture of GLMs}. An empirical analysis based on simulated data will be presented in Example \ref{Estimates comparison between the MLCM and the constrained GLGCWM} of Section \ref{sec:numerical studies}.
\item If the marginal distributions $p(\bx; \bmu_g, \bSigma_g)$  in \eqref{eq:CWM base} are Gaussian with the same covariance matrices, i.e. $p(\bx; \bmu_g, \bSigma_g)=\phi_d(\bx; \bmu_g, \bSigma)$ and the mixing weights are equal, i.e. $\pi_g=1/G$ ($g=1, \ldots, G)$,
then the estimates $\{\bbeta_g, \lambda_g,  \; g=1, \ldots, G\}$ in \eqref{eq:CWM base} and  \eqref{eq:finite mixture of GLMs with concomitant} are the same according to Proposition \ref{CWM->FMRC}; moreover,  the posterior probability \eqref{eq:CWM posteriors} reduces to \eqref{eq:posterior from finite mixture of GLMs with multinomial concomitant}.
\end{itemize}

\section{The EM algorithm for the Generalized Linear CWM}
\label{sec:EM} 

In this section, we present the main steps of the EM algorithm for parameter estimation of the Generalized Linear CWM in \eqref{eq:Generalized Linear Gaussian CWM}.
In this case, the three terms of the complete-data log-likelihood function \eqref{llikCWM} are given by:

\begin{align*}
\cL_{1c} (\bxi) & = \sum_{n=1}^N \sum_{g=1}^G  z_{ng} \left[ \frac{y_n \eta(\bx_n; \bbeta_g) - b(\eta(\bx_n; \bbeta_g))}{a(\lambda_g )} + c(y,\lambda_g)   \right] \\
\cL_{2c} (\bpsi) & = \sum_{n=1}^N \sum_{g=1}^G  z_{ng} \frac{1}{2} \left[- p \ln 2\pi - \ln |\bSigma_g| - (\bx_n-\bmu_g)' \bSigma_g^{-1} (\bx_n-\bmu_g)\right] \\
\cL_{3c} (\pi) & = \sum_{n=1}^N \sum_{g=1}^G  z_{ng} [\ln \pi_g].
\end{align*}

The {\it E-step} on the $(k+1)$-th iteration of the EM algorithm requires the calculation of the conditional expectation of the complete-data log-likelihood function $\cL_c(\btheta; \sbX, \by)$ in \eqref{llikCWM}, say $Q(\btheta,\btheta^{(k)})$, evaluated using the current fit $\btheta^{(k)}$ for $\btheta$. Since $\cL_c(\btheta; \sbX, \by)$ is linear in the unobservable data $z_{ng}$, this means calculating the current conditional expectation of $Z_{ng}$ given $\sbX$ and $\by$, where $Z_{ng}$ is the random variable corresponding to $z_{ng}$, that is
\begin{align}
Q(\btheta,\btheta^{(k)}) & = \EE_{\theta^{(k)}} \{ \cL_c(\btheta;\sbX, \by)\}  \notag \\
& = \sum_{n=1}^N \sum_{g=1}^G  \EE_{\theta^{(k)}} \{Z_{ng}|\bx_n, y_n\} [ Q_{1}(\bbeta_g, \lambda_g;\btheta^{(k)}) + Q_{2}(\bmu_g, \bSigma_g;\btheta^{(k)}) + \ln \pi_g]  \notag \\
& = \sum_{n=1}^N \sum_{g=1}^G  \tau_{ng}^{(k)} [ Q_{1}(\bbeta_g, \lambda_g;\btheta^{(k)}) + Q_{2}(\bmu_g, \bSigma_g;\btheta^{(k)}) + \ln \pi_g], \label{Qbtheta(k)}
\end{align}
where 
$Q_{1}(\bbeta_g, \lambda_g;\btheta^{(k)})$ and $Q_{2}(\bmu_g, \bSigma_g;\btheta^{(k)})$ depend on the functional form
of densities $p(y_n|\bx_n; \bbeta_g, \lambda_g)$ and $ p(\bx_n; \bmu_g, \bSigma_g)$, respectively: 
\begin{align} 
Q_{1}(\bbeta_g, \lambda_g;\btheta^{(k)}) &=    \frac{y_n \eta(\bx_n; \bbeta_g) - b(\eta(\bx_n; \bbeta_g))}{a(\lambda_g )} + c(y_n,\lambda_g)   ,
\label{Q1bxi;theta}\\
Q_{2}(\bmu_g, \bSigma_g;\btheta^{(k)}) &=  \frac{1}{2} \left[- p \ln 2\pi - \ln |\bSigma_g| - (\bx_n-\bmu_g)' \bSigma_g^{-1} (\bx_n-\bmu_g)\right].
\label{Q2bxi;theta}
\end{align}
The {\em M-step} on the $(k+1)$-th iteration of the EM algorithm requires the maximization of the conditional expectation of the complete-data log-likelihood $Q(\btheta,\btheta^{(k)})$ with respect to $\btheta$.
The maximization of the quantity in \eqref{Q1bxi;theta} is equivalent to the maximization problem of the generalized linear model for the complete data, except that each observation $y_{n}$ contributes to the log-likelihood for each group $g$ with a known weight $\tau_{ng}^{(k)}$. The stationary equations are obtained by equating the first order partial derivatives of $\sum_{n=1}^N \sum_{g=1}^G  \tau_{ng}^{(k)} Q_{1}(\bbeta_g, \lambda_g;\btheta^{(k)})$
to zero, that is
\begin{align*}
& \frac{\partial}{\partial \bbeta_g} \sum_{n=1}^N \sum_{g=1}^G  \tau_{ng}^{(k)}  \frac{y_n \eta(\bx_n; \bbeta_g) - b(\eta(\bx_n; \bbeta_g))}{a(\lambda_g)}=0 \\
& \frac{\partial}{\partial \lambda_g} \sum_{n=1}^N \sum_{g=1}^G  \tau_{ng}^{(k)}  \frac{y_n \eta(\bx_n; \bbeta_g) - b(\eta(\bx_n; \bbeta_g))}{a(\lambda_g)}=0
\end{align*}
which after some algebra yield
\begin{align*}
& \sum_{n=1}^N \sum_{g=1}^G  \tau_{ng}^{(k)} \frac{y_n - b'(\eta(\bx_n; \bbeta_g))}{a(\lambda_g)} x_{npg} \frac{\partial}{\partial \eta_g} \mu_g \frac{1}{V_{ng}}=0 \\
& \sum_{n=1}^N \sum_{g=1}^G  \tau_{ng}^{(k)} \frac{-[y_n \eta(\bx_n; \bbeta_g) - b(\eta(\bx_n; \bbeta_g))]}{[a(\lambda_g )]^2} \frac{\partial}{\partial \lambda_g} a(\lambda_g ) + \frac{\partial}{\partial \lambda_g} c(y_n,\lambda_g)=0.
\end{align*}
Maximization can be obtained by the iterative reweighted least-squares procedure by \citet{Nel:Wed:1972} for ML estimation of generalized linear models, with each observation $y_n$ weighted additionally with $\tau_{ng}^{(k)}$. 

For the maximization of the quantity in \eqref{Q2bxi;theta}, the solutions for the weights $\pi_{g}^{(k+1)}$ and the parameters estimates $\bmu_g^{(k+1)}$, $\bSigma_g^{(k+1)}$  of the local  densities are given in closed form, that is: 
\begin{align*}
\pi_{g}^{(k+1)} &= \frac{1}{N} \sum_{n=1}^N \tau_{ng}^{(k)} \\
\bmu_g^{(k+1)} &= \frac{\sum_{n=1}^N \tau_{ng}^{(k)}  \bx_n}{\sum_{n=1}^N \tau_{ng}^{(k)} u_{ng}^{(k)}  } \\
\bSigma_g^{(k+1)} &= \frac{\sum_{n=1}^N \tau_{ng}^{(k)}  (\bx_n - \bmu_g^{(k+1)})(\bx_n - \bmu_g^{(k+1)})'}{\sum_{n=1}^N \tau_{ng}^{(k)} }, 
\end{align*}
see e.g. \citet{McLa:Peel:fini:2000}.
In the rest of the section, we present two computational issues.

\paragraph{Algorithm initialization}

The algorithm has been initialized by assigning an initial classification of the units, that is by specifying a value for $\boldsymbol{z}_n^{(0)}$, $n=1,\ldots,N$ \citep[see, e.g.,][]{McLa:Peel:fini:2000}.   

\paragraph{Convergence criterion}

The convergence criterion is based on the 
Aitken acceleration procedure \citep{Aitk:OnBe:1926} which is used to estimate the asymptotic maximum of the log-likelihood at each iteration of the EM algorithm. 
Based on this estimate, a decision can be made regarding whether or not the algorithm has reached convergence;
that is, whether or not the log-likelihood is sufficiently close to its estimated asymptotic value. 
The Aitken acceleration at iteration $k$ is given by
\begin{displaymath}
	a^{(k)}=\frac{l^{(k+1)}-l^{(k)}}{l^{(k)}-l^{(k-1)}},
\end{displaymath}
where $l^{(k+1)}$, $l^{(k)}$, and $l^{(k-1)}$ are the log-likelihood values from iterations $k+1$, $k$, and $k-1$, respectively. 
Then, the asymptotic estimate of the log-likelihood at iteration $k + 1$ is given by
\begin{displaymath}	l_{\infty}^{(k+1)}=l^{(k)}+\frac{1}{1-a^{(k)}}(l^{(k+1)}-l^{(k)}),
\end{displaymath}
see \citet{Bohn:Diet:Scha:Schl:Lind:TheD:1994}.
In the analyses in Section~\ref{sec:numerical studies}, the algorithms stopped when $l_{\infty}^{(k+1)}-l^{(k)}<\epsilon$, with $\epsilon=0.05$.

\section{Performance evaluation}
\label{sec:Performance evaluation}
In order to evaluate the performance of the models introduced in the paper, some different indices will be taken into account. They are classified according to the type of the available data. In general, model will be selected according to the BIC, see \ref{sec:model selection}. 
Moreover, when data are simulated and the data labeling is known  we can use indices which compare the true partition with that arising from the application of a particular model (like the misclassification error and indices described in Section~\ref{subsec:Adjusted Rand index}).
On the contrary, when we do not know the true labels, we limit our attention to goodness-of-fit indices (see Section~\ref{subsec:The generalized weighted Pearson chi-square statistic} and Section~\ref{subsec:The generalized scaled deviance}).
 


\subsection{Model selection and performance}\label{sec:model selection}
In our numerical studies, we considered the Bayesian information criterion (BIC) \citet{Schw:Esti:1978}:
\begin{displaymath}
\text{BIC}= 2 l(\widehat{\btheta}) - m \ln N.
\end{displaymath}
where $\widehat{\btheta}$ is the ML estimate of $\btheta$, $l(\widehat{\btheta})$ is the maximized observed-data log-likelihood, and $m$ is the overall number of free parameters in the model.

\subsection{The Rand index and the adjusted Rand index}
\label{subsec:Adjusted Rand index}

When the true classification is known, often  the adjusted Rand index \citep[ARI;][]{Hube:Arab:Comp:1985} is considered
 as a measure of class agreement. 
The original Rand index \citep[RI;][]{Rand:Obje:1971} can be expressed
as
\begin{displaymath}
\text{RI}=\frac{\text{number of agreements}}{\text{number of agreements + number of disagreements}},	
\end{displaymath}
where the number of agreements (observations that should be in the same group and are, plus those that should not be in the same group and are not) and the number of disagreements are based on pairwise comparisons.
The RI assumes values between 0 and 1, where 0 indicates no pairwise agreement between the MAP classification and true group membership and 1 indicates perfect agreement.

One criticism of the RI is that its expected value is greater than 0, making smaller values difficult to interpret.
The ARI corrects the RI for chance by allowing for the possibility that classification performed randomly will correctly classify some observations, see \citet{Hube:Arab:Comp:1985}.
The ARI can be expressed as
\begin{displaymath}
\text{ARI}=\frac{\text{RI} - \EE(\text{RI})}{\max(\text{RI}) - \EE(\text{RI})}, 
\end{displaymath}
where the expected value $\EE(\text{RI})$ of the Rand Index is computed considering all pairs of distinct partitions picked at random, subject
to having the original number of classes and objects in each. 
Thus, the ARI has an expected value of 0 and perfect classification would result in a value equal to 1.

\subsection{The index of conditional goodness-of-fit}
\label{subsec:The generalized weighted Pearson chi-square statistic}

We  introduce a descriptive measure of goodness-of-fit which is directly related to the generalized Pearson $\chi^2$ statistic commonly used for generalized linear models.
In detail, our index of conditional goodness-of-fit (CGOF) is defined as 
\begin{equation}
\text{CGOF} = \frac{\chi^2_w}{N}= \frac{1}{N}\sum_{n=1}^N\sum_{g=1}^G\widehat{z}_{ng}\frac{\left[y_n-\mathbb{E}(Y|\bx_n,\Omega_g)\right]^2}{\mathbb{V}(Y|\bx_n,\Omega_g)}. 
\label{eq:CGOF}
\end{equation}
As a special case, when $G=1$, $\chi^2_w$ corresponds to the classical generalized Pearson $\chi^2$ statistic (see, e.g., \citealp[][p.~34]{McCullaghNelder:GLM:2000} and \citealp[][p.~46]{Olss:Gene:2006}).
In \eqref{eq:CGOF} the quantity $\chi^2_w$ is divided by $N$ in order to remove the impact of the sample size.
Furthermore, the division by the conditional variance $\mathbb{V}(Y|\bx_n,\Omega_g)$ makes the squared residuals $\left[y_n-\mathbb{E}(Y|\bx_n,\Omega_g)\right]^2$ comparable between groups.

 
\subsection{An index of goodness-of-fit based on scaled deviance}
\label{subsec:The generalized scaled deviance}

Another goodness-of-fit criterion consists of an extension of the traditional scaled deviance for GLM (see \citealp[p.34]{McCullaghNelder:GLM:2000}), which is defined as:
\begin{equation}
SD=D^*(\by;\bmu)=2l(\by;\by)-2l(\bmu,\by),
\label{eq:scdev1}
\end{equation}
that is the difference between the maximum likelihood achievable in a full model and that achieved by the model under investigation.
Note that
\begin{equation}
D^*(\by;\hat{\bmu}) = D(\by;\hat{\bmu})/\phi,
\label{eq:scdev2}
\end{equation}
where $D(y;\hat{y})$ is the deviance for the current model, so that the scaled deviance is the deviance expressed as a multiple of the dispersion parameter.

Given the forms of the deviances for Binomial and Poisson distributions, respectively (see \citealp[p.34]{McCullaghNelder:GLM:2000}), the corresponding measures for GLGCWM, called generalized deviances, can be defined as:
\begin{align*}
& GD_{\rm Bin} = 2\sum_{n=1}^N\sum_{g=1}^G\widehat{z}_{ng}
\left\{y_{n} \log \left(\frac{y_n}{\mathbb{E}(Y|\bx_n,\Omega_g)}\right) 
+ (K - y_n) \log \left(\frac{M - y_n}{M - \mathbb{E}(Y|\bx_n,\Omega_g)}\right)\right\} \\
& GD_{\rm Poi} = 2\sum_{n=1}^N\sum_{g=1}^G\widehat{z}_{ng}
\left\{y_n \log \left(\frac{y_n}{\mathbb{E}(Y|\bx_n,\Omega_g)}\right) - \left[y_n - \mathbb{E}(Y|\bx_n,\Omega_g)\right]\right\}.
\end{align*}
Analogous forms are obtained for the generalized scaled deviances, by dividing the general deviances for the dispersion parameter, which is $1/M$ for the Binomial distribution and $1$ for the Poisson distribution, respectively, that is
\begin{align*}
& GSD_{\rm Bin}=GD_{\rm Bin}/M\\
& GSD_{\rm Poi}=GD_{\rm Poi}.
\end{align*}
 
%

\section{Numerical studies}
\label{sec:numerical studies}

In this section we illustrate some features of the models we proposed above, on
the ground of numerical studies based on both artificial and real data. We aim also at a comparison with the competitive models, that is FMGLM and FMGLMC.
Code for all of the analyses presented herein was written in the \texttt{R} computing environment \citep{R}.
We remark that the parameter estimation of FMGLM and FMGLMC has been carried out by means of the \texttt{R}-package \texttt{flexmix},  see \citet{Leis:Flex:2004},  \citet{Grun:Leis:Flex:2008}.

\subsection{Simulated data}
\label{sec:simdata}

\begin{Ex}[Estimates comparison between FMGLM and constrained GLGCWM]
\label{Estimates comparison between the MLCM and the constrained GLGCWM}{\rm

In Section~\ref{sec:Parameter estimation} we stated that FMGLM can be considered as  nested models of GLGCWM. 
In particular, we showed that if  $\bmu_g, \bSigma_g=\bmu, \bSigma$ ($g=1,\ldots,G$), then the conditional distributions in GLGCWM and FMGLM have the same estimates of $\left\{\bbeta_g, \lambda_g,\pi_g;g=1,\ldots,G\right\}$. 
Thus, first numerical simulations concern the comparison between such estimates in data modeling according to either Poisson Gaussian CWM and mixture of Poisson regressions when the marginal distributions do not depend on the $g$th group (in the following, this case  will be referred to as 
{\em constrained Poisson Gaussian CWM}). 
Here we considered $G=2$ groups.

The data have been generated as follows. 
As for the  marginal distributions, we considered three cases: $X \sim N(5,0.8)$, $X \sim \text{Unif}(4.4,5.5)$, and  $X \sim \text{Unif}(4,5)$.
In particular, we remark that two out three cases concern  non Gaussian marginal distributions. 
As for the conditional distributions, data have been generated according to a Poisson distribution \eqref{eq:Poisson Y} with the following parameters: $\beta_{01}=1$ and $\beta_{11}=0.2$ in the first group, $\beta_{02}=0$ and $\beta_{12}=0.6$ in the second group. 
For each combination of the parameters, we have generated 120 random samples with sample sizes $N_1=250$ and $N_2=350$.

In each replication, once both the models were fitted to the generated data, the following index was computed to evaluate the discrepancy
\begin{equation}
	\sum_{i=1}^2\sum_{j=0}^1\left|\widehat{\beta}_{ij}^{\text{GLGCWM}}-\widehat{\beta}_{ij}^{\text{FMGLM}}\right|/4,
	\label{eq:L1 distance}
\end{equation} 
where $\widehat{\beta}_{ij}^{\text{GLGCWM}}$ and $\widehat{\beta}_{ij}^{\text{FMGLM}}$ are the ML estimates of $\beta_{ij}$ for the Poisson GCWM and the Poisson GFMR, respectively.
For each of the three considered $X$-distributions, the average values, with respect to the 120 replications, of the index in \eqref{eq:L1 distance} were respectively: 0.00131 when $X \sim N(5,0.8)$, 0.00386 when $X \sim \text{Unif}(4.4,5.5)$, and 0.00259 when $X \sim \text{Unif}(4,5)$.
These results underline as the constrained Poisson GCWM reveals to be a very good approximation for the Poisson GFMR, regardless from the distribution of $X$.
}\end{Ex}

\begin{Ex}[Data from a Binomial Gaussian CWM]
\label{ex:Binomial-based models}{\rm

The $N=600$ artificial bivariate data of this example are referred to $G=2$ groups of size $N_1=250$ and $N_2=350$.
They are randomly generated from a Binomial Gaussian CWM, with $M=30$, with parameters given in the first row of \tablename~\ref{tab:Bin_parameters Poisson CWM} (here ``true parameters'' means the parameters we used for data generation).
\begin{table}[!ht]
\caption{Example \ref{ex:Binomial-based models}. 
Parameters and estimates according to the Binomial Gaussian CWM.
Standard errors are given in round brackets}
\label{tab:Bin_parameters Poisson CWM}
\centering
\resizebox*{1\textwidth}{!}{
\begin{tabular}{lcccccccccc}
\hline
& $\pi_1$ & $\pi_2$ & $\mu_1$ & $\mu_2$ & $\sigma_1$ & $\sigma_2$ & $\beta_{01}$ & $\beta_{02}$ & $\beta_{11}$ & $\beta_{12}$\\ 
\hline
true parameters	&	0.417	&	0.583	&	2.000	&	-2.000	&	1.000	&	1.000	&	0.000	&	0.000	&	1.000	&	1.000	\\[2mm]
estimates	  &	0.427	&	0.573	&	1.933	&	-1.986	&	0.996	&	0.936	&	0.082	&	-0.059	&	0.969	&	0.941	\\
					  &	(0.028)	&	(0.032)	&	(0.074)	&	 (0.059)	&	(0.114)	&	(0.088)	&	(0.081)	&	(0.071)	&	(0.047)	&	(0.040)	\\
\hline
\end{tabular}
}
\end{table}
 Figure \ref{fig:Bin_artificial-CWMscatter} displays the scatterplot of the data and  the fitted models.
\begin{figure}[!ht]
\centering
\subfigure[TRUE\label{fig:Bin_scatterTRUE}]
{\includegraphics[width=0.49\textwidth]{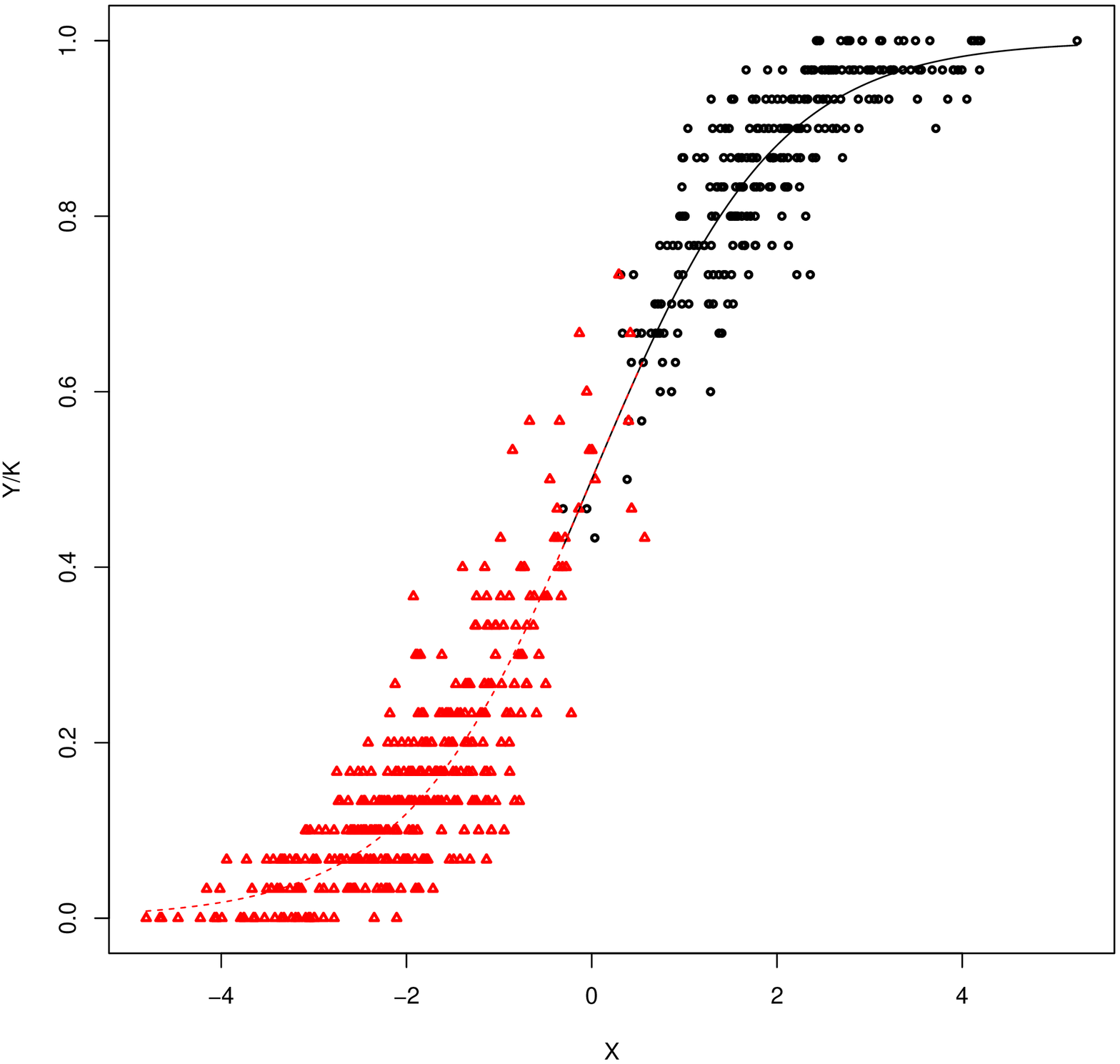}}
\subfigure[Binomial FMR\label{fig:Bin_scatterFMR}]
{\includegraphics[width=0.49\textwidth]{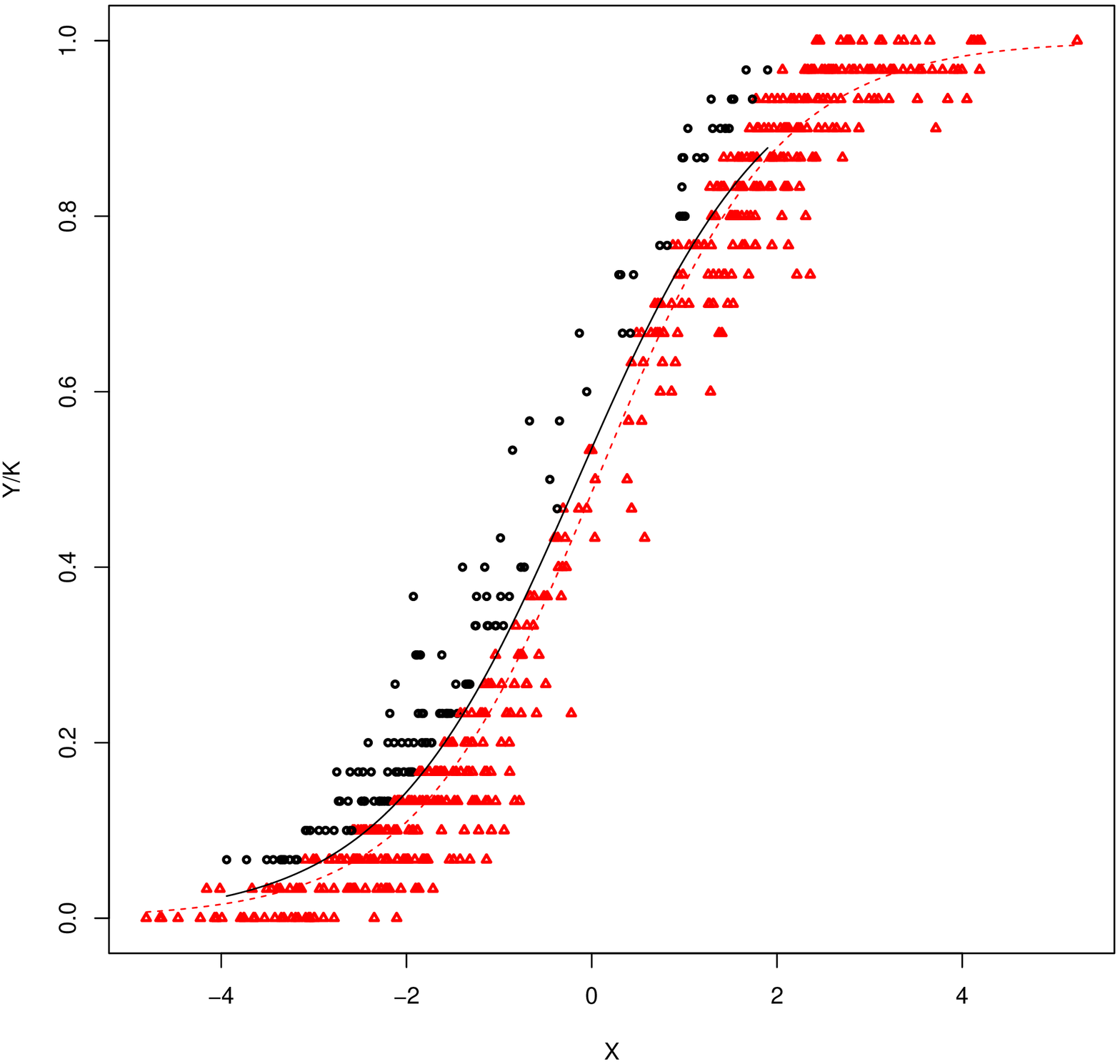}}
\subfigure[Binomial FMRC\label{fig:Bin_scatterFMRC}]
{\includegraphics[width=0.49\textwidth]{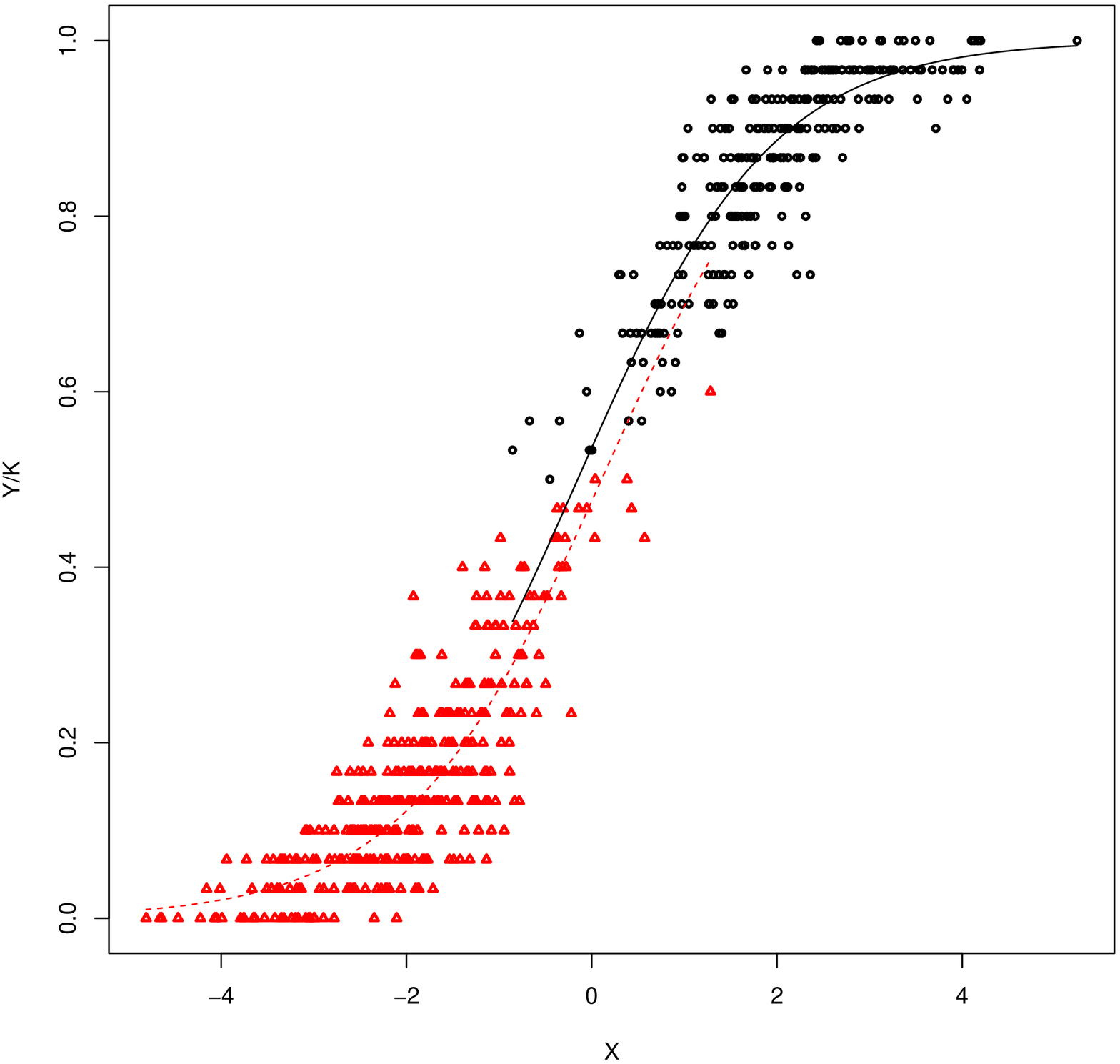}}
\subfigure[Binomial CWM\label{fig:Bin_scatterCWM}]
{\includegraphics[width=0.49\textwidth]{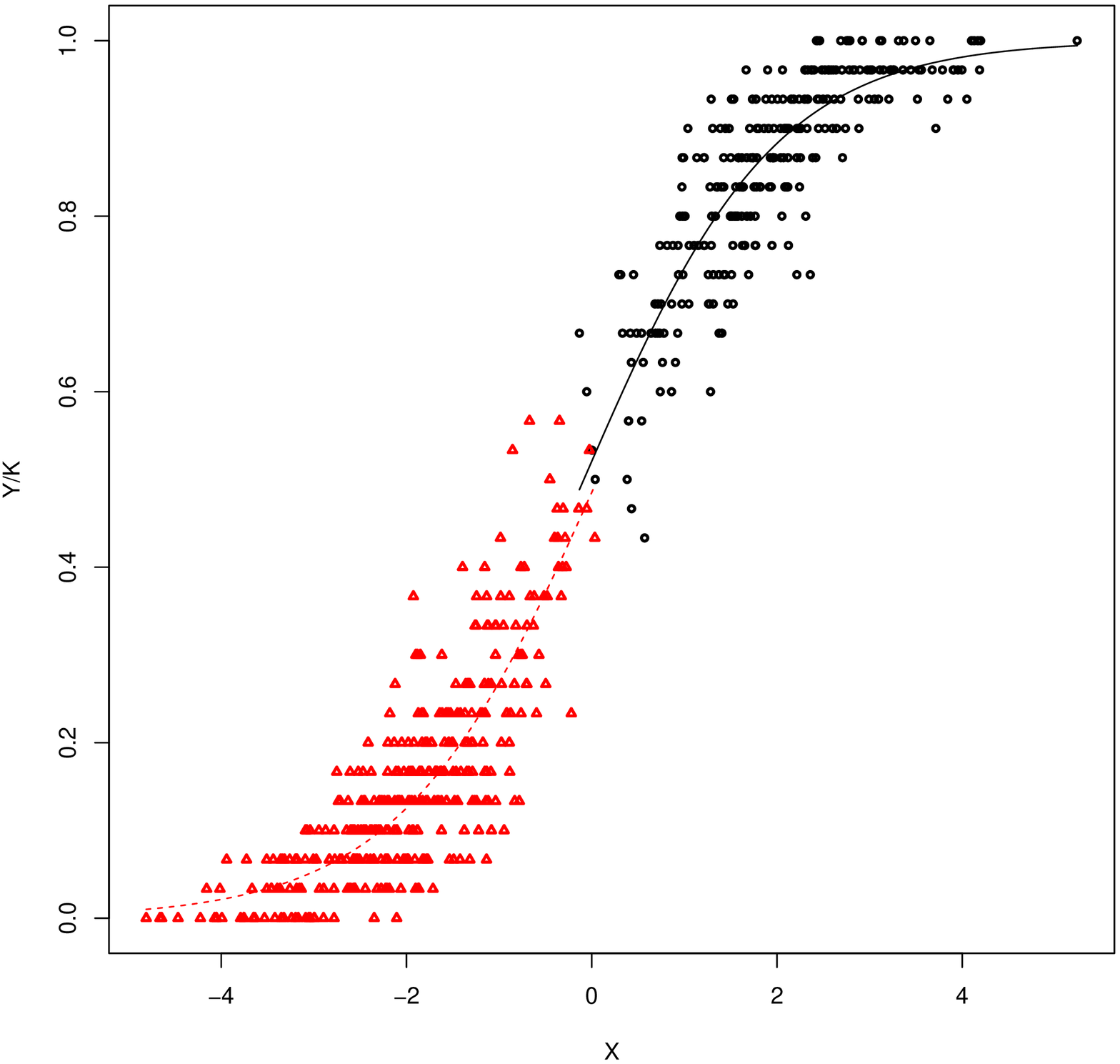}}
\caption{Example \ref{ex:Binomial-based models}. 
Scatter plots and Binomial-based models.
}
\label{fig:Bin_artificial-CWMscatter}
\end{figure}
\figurename~\ref{fig:BCWM density} displays the joint density from a Binomial CWM for different values of $y$.
\begin{figure}[!ht]
\centering
\includegraphics[width=10cm]{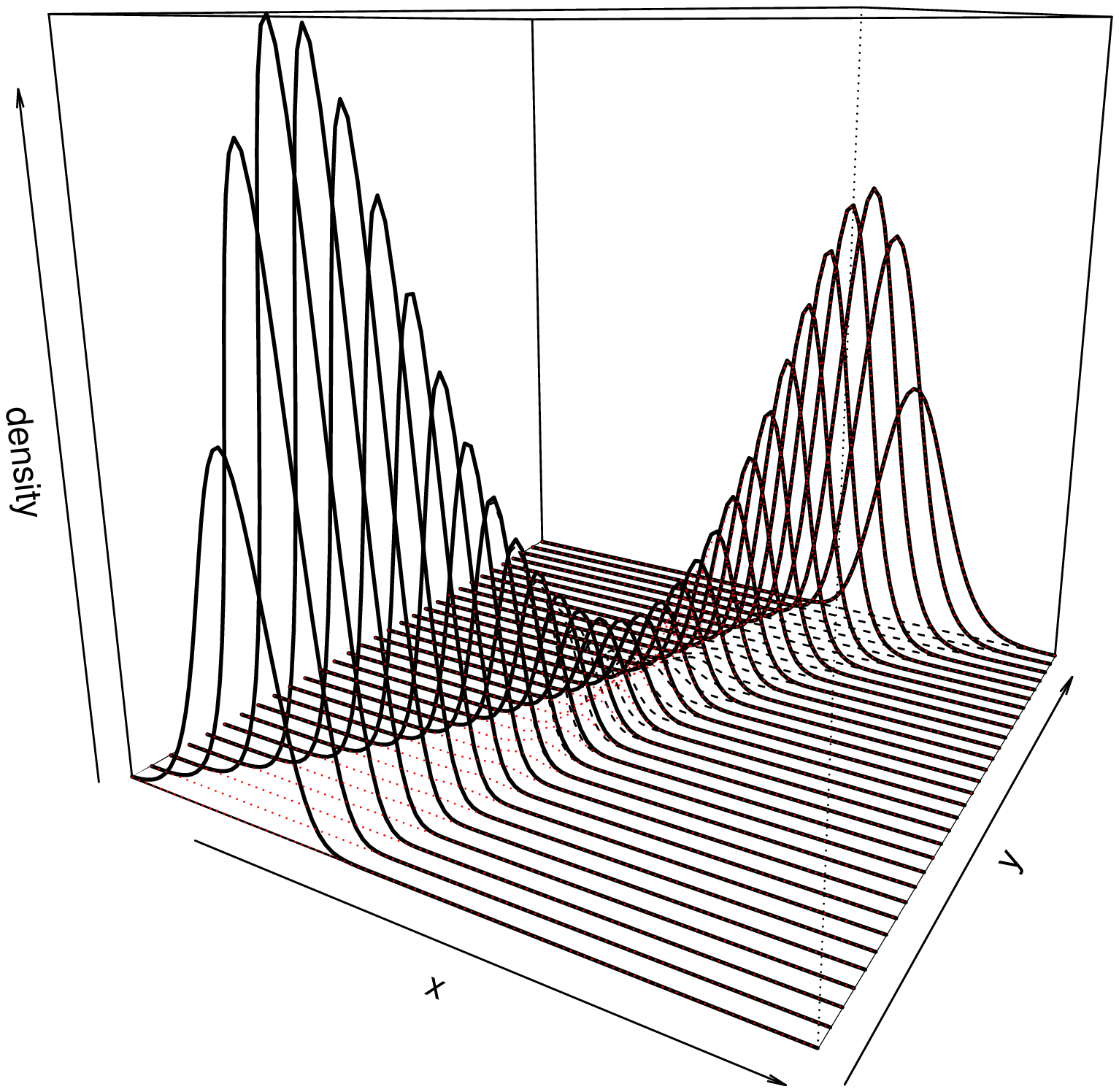} 
\caption{Example \ref{ex:Binomial-based models}. 
Joint density from the fitted Binomial CWM.}
\label{fig:BCWM density}
\end{figure}
In \tablename~\ref{tab:Bin_Clustering performance} we present the ARIs and the misclassification errors we obtained in the three data modeling.
Here, it is possible to see as the Binomial Gaussian CWM outperforms the other approaches; in particular, we remark that the Binomial MR is not able to capture the underlying group-structure of the data.  
\begin{table}[!ht]
\caption{\it Example \ref{ex:Binomial-based models}.  
Clustering performance of the fitted Binomial-based models}
\label{tab:Bin_Clustering performance}
\centering
\begin{tabular}{rrrr}
\hline
	&	FMR	&	FMRC	&	CWM	\\
\hline
misclassError	&	41.67\%	&	2.67\%	&	2.00\%	\\
ARI	&	-0.00078	&	0.89595	&	0.92143	\\
\hline
\end{tabular}
\end{table}
}\end{Ex}

\begin{Ex}[Data from a Binomial Gaussian CWM.]
\label{ex:Binomial-based models2}{\rm

The $N=250$ artificial bivariate data of this example are referred to $G=2$ groups of size $N_1=100$ and $N_2=150$.
They are randomly generated from a Binomial Gaussian CWM, with $M=30$, with parameters given in the first row of \tablename~\ref{tab:Bin_parameters Poisson CWM 2}.
\begin{table}[!ht]
\caption{
Example~\ref{ex:Binomial-based models2}. 
Parameters and estimates according to the Binomial Gaussian CWM.
Standard errors are given in round brackets
}
\label{tab:Bin_parameters Poisson CWM 2}
\centering
\resizebox*{1\textwidth}{!}{
\begin{tabular}{lcccccccccc}
\hline
& $\pi_1$ & $\pi_2$ & $\mu_1$ & $\mu_2$ & $\sigma_1$ & $\sigma_2$ & $\beta_{01}$ & $\beta_{02}$ & $\beta_{11}$ & $\beta_{12}$\\ 
\hline
true parameters	&	0.400	&	0.600	&	0.000	&	 0.000	&	1.000	&	16.000	&	0.000	&	0.000	&	2.000	&	0.500	\\[2mm]
estimates	  &	0.388	&	0.612	&	0.069	&	-0.277	&	0.969	&	14.252	&	0.081	&	-0.059	&	1.952	&	0.494	\\
					  &	(0.044)	&	(0.053)	&	(0.106)	&	 (0.305)	&	(0.159)	&	(1.688)	&	(0.058)	&	(0.042)	&	(0.089)	&	(0.015)	\\
\hline
\end{tabular}
}
\end{table}
\figurename~\ref{fig:Bin_artificial-CWMscatter2} displays the scatterplot of the data and  the fitted models.
\begin{figure}[!ht]
\centering
\subfigure[TRUE\label{fig:Bin_scatterTRUE2}]
{\includegraphics[width=0.49\textwidth]{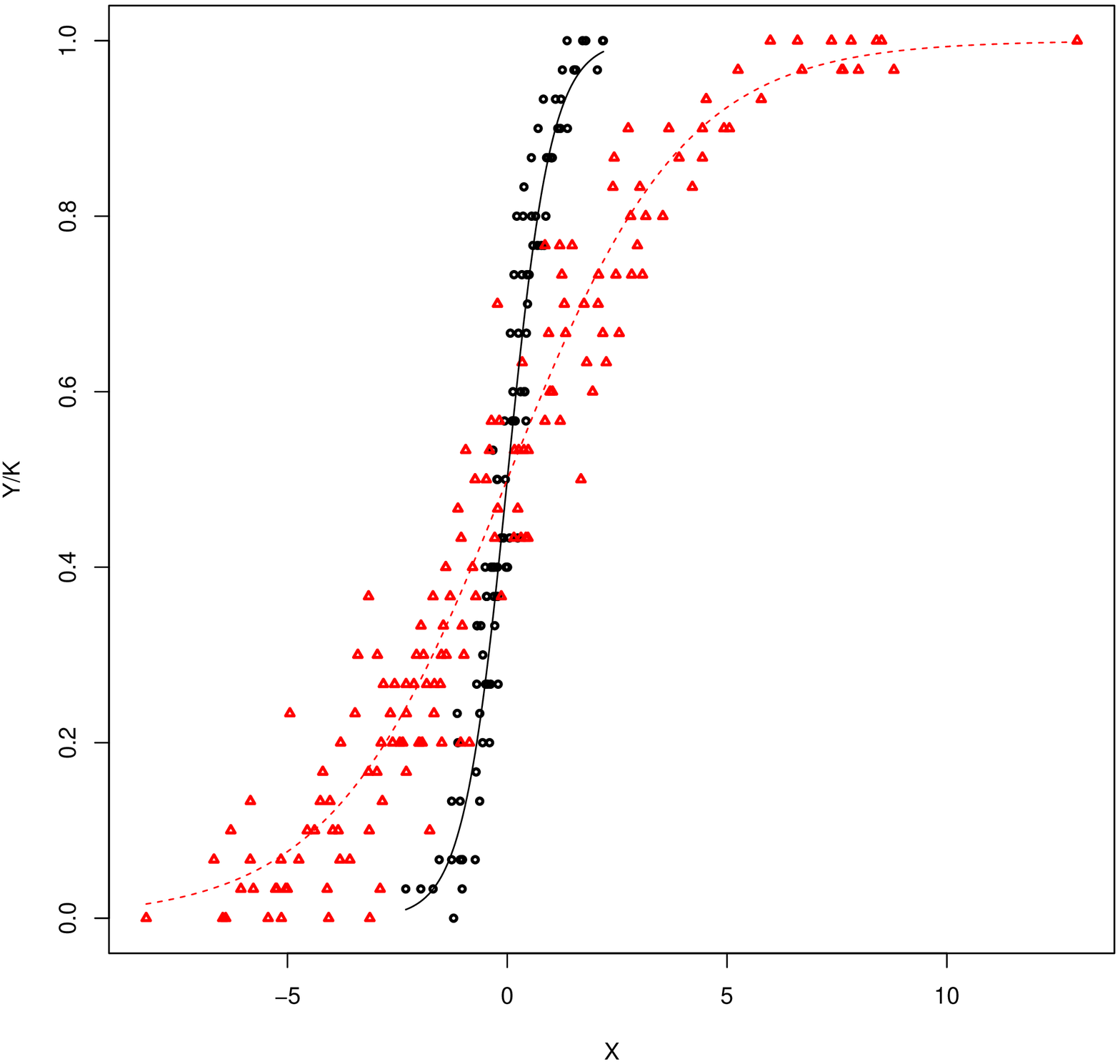}}
\subfigure[Binomial FMR\label{fig:Bin_scatterFMR2}]
{\includegraphics[width=0.49\textwidth]{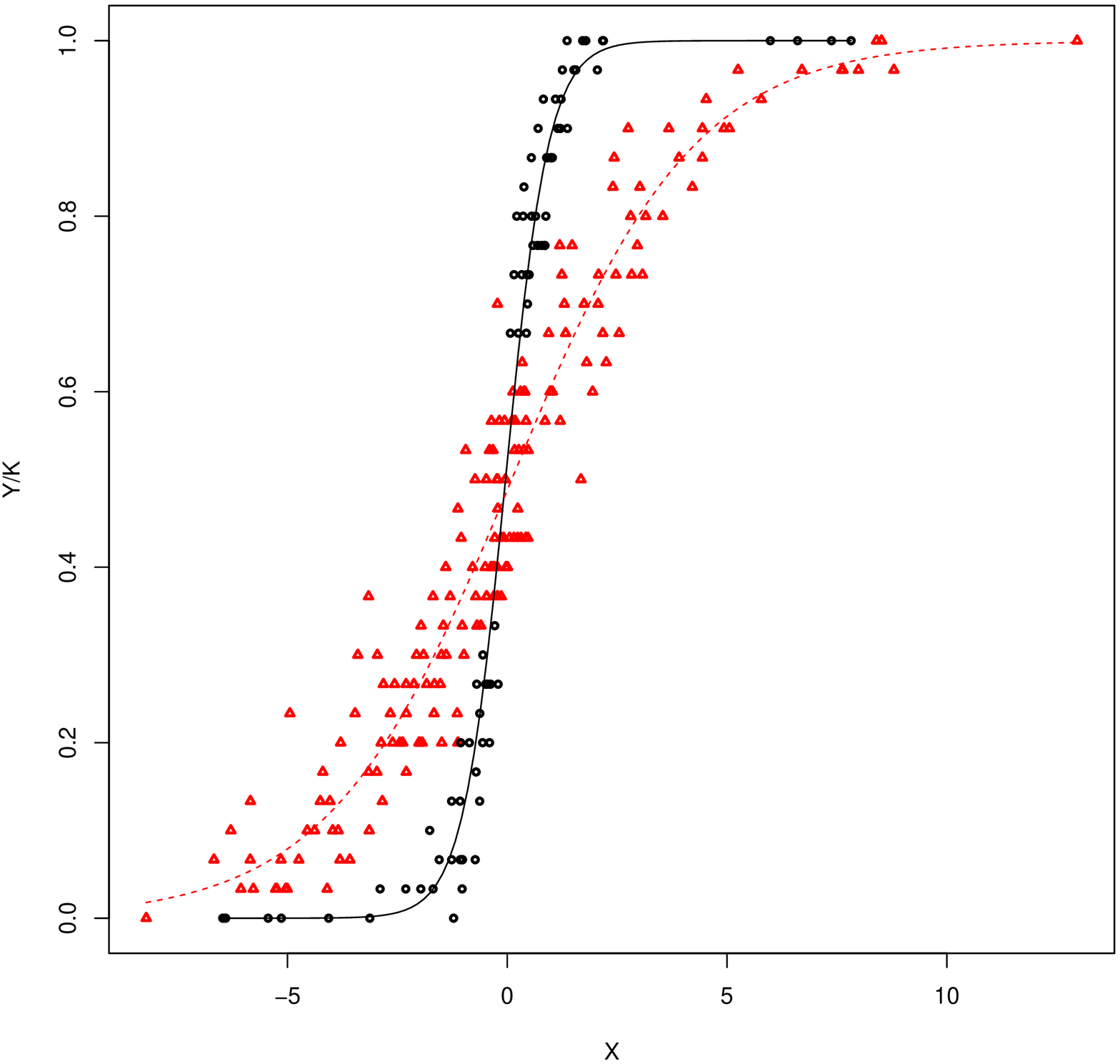}}
\subfigure[Binomial FMRC\label{fig:Bin_scatterFMRC2}]
{\includegraphics[width=0.49\textwidth]{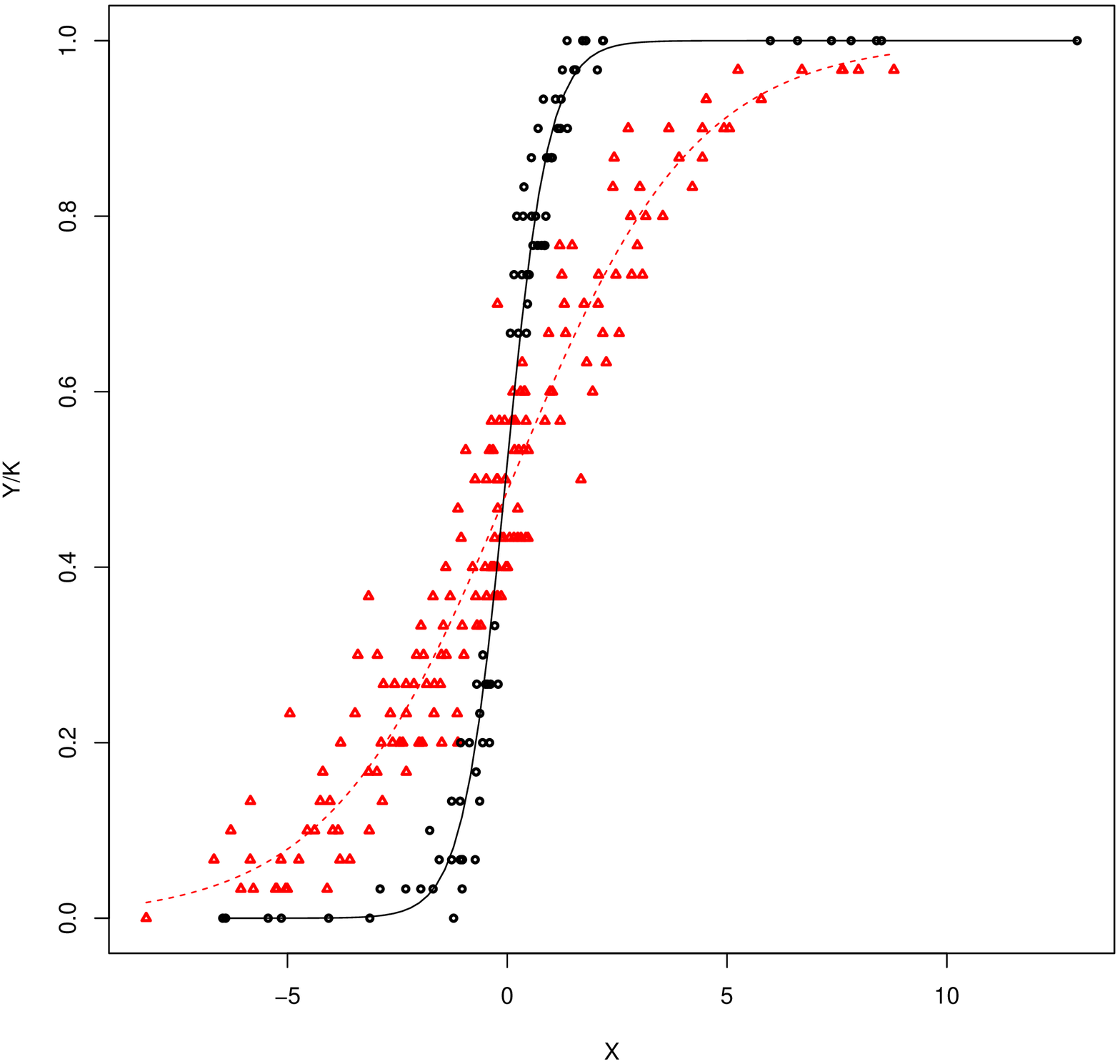}}
\subfigure[Binomial CWM\label{fig:Bin_scatterCWM2}]
{\includegraphics[width=0.49\textwidth]{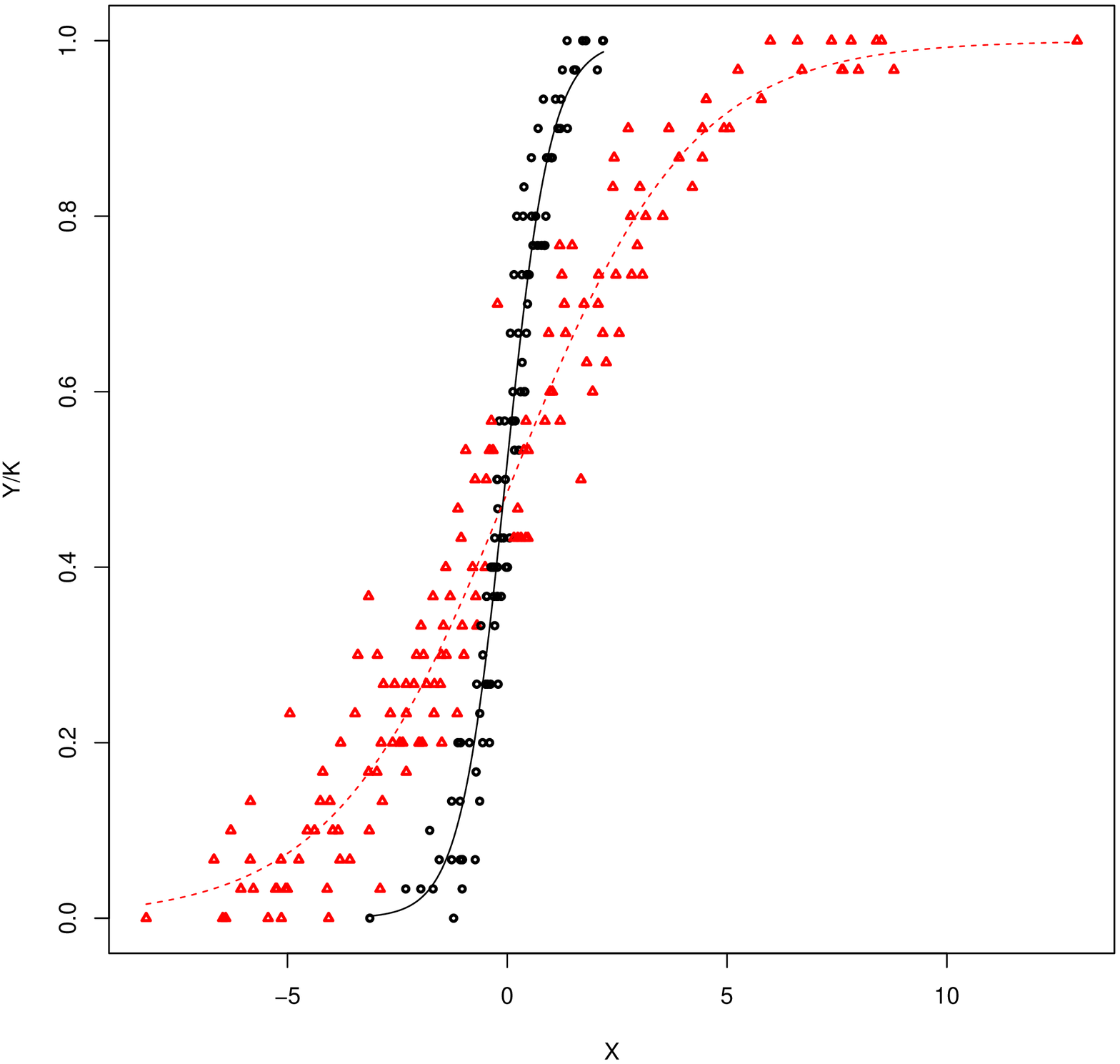}}
\caption{Example \ref{ex:Binomial-based models2}.
Scatter plots and Binomial-based models.
}
\label{fig:Bin_artificial-CWMscatter2}
\end{figure}
In \tablename~\ref{tab:Bin_Clustering performance2} we list the ARIs and the misclassification errors we obtained in the three data modeling.
Like in Example~\ref{ex:Binomial-based models2}, the Binomial Gaussian CWM outperforms the other approaches but, differently from the previous case, here  the Binomial Gaussian CWM clearly outperforms the FMRC.
Moreover, it is possible to see like the clustering results for the FMRC are the worse than those obtained via FMR. 
\begin{table}[!ht]
\caption{\it Example~\ref{ex:Binomial-based models2}.  
Clustering performance of the fitted Binomial-based models}
\label{tab:Bin_Clustering performance2}
\centering
\begin{tabular}{rrrr}
\hline
	&	FMR	&	FMRC	&	CWM	\\
\hline
misclassError	&	19.60\%	&	20.80\%	&	7.60\%	\\
ARI	&	0.36436	&	0.33591	&	0.71775	\\
\hline
\end{tabular}
\end{table}
}\end{Ex}

\begin{Ex}[Data from a Poisson GCWM.]
\label{ex:Data from a Poisson GCWM}{\rm 
The $N=400$ artificial bivariate data of this example are referred to $G=2$ groups of size $N_1=150$ and $N_2=250$, respectively.
They are randomly generated from a Poisson CWM with parameters specified in the first row of \tablename~\ref{tab:parameters Poisson CWM}.
\begin{table}[!ht]
\caption{\it Example \ref{ex:Data from a Poisson GCWM}. Parameters  and estimates according to Poisson CWM.
Standard errors are given in round brackets.}
\label{tab:parameters Poisson CWM}
\centering
\resizebox*{1\textwidth}{!}{
\begin{tabular}{lcccccccccc}
\hline
& $\pi_1$ & $\pi_2$ & $\mu_1$ & $\mu_2$ & $\sigma_1$ & $\sigma_2$ & $\beta_{01}$ & $\beta_{02}$ & $\beta_{11}$ & $\beta_{12}$\\ 
\hline
true parameters      &	0.375	&	0.625	&	 0.000	&	5.000	&	1.500	&	0.800	&	1.000	&	0.000	&	0.200	&	0.500\\[2mm]
estimates        &	0.370	&	0.630	&	-0.061	&	5.016	&	1.291	&	0.817	&	0.985	&	0.005	&	0.203	&	0.497\\
                 &	(0.031)	&	(0.040)	&	 (0.100)	&	(0.060)	&	(0.174)	&	(0.082)	&	(0.051)	&	(0.103)	&	(0.047)	&	(0.019)\\
\hline
\end{tabular}
}
\end{table}
Figure \ref{fig:artificial-CWMscatter} displays the scatterplot of the data and the fitted Poisson-based models described in the paper.
\begin{figure}[!ht]
\centering
\subfigure[TRUE\label{fig:scatterTRUE}]
{\includegraphics[width=0.49\textwidth]{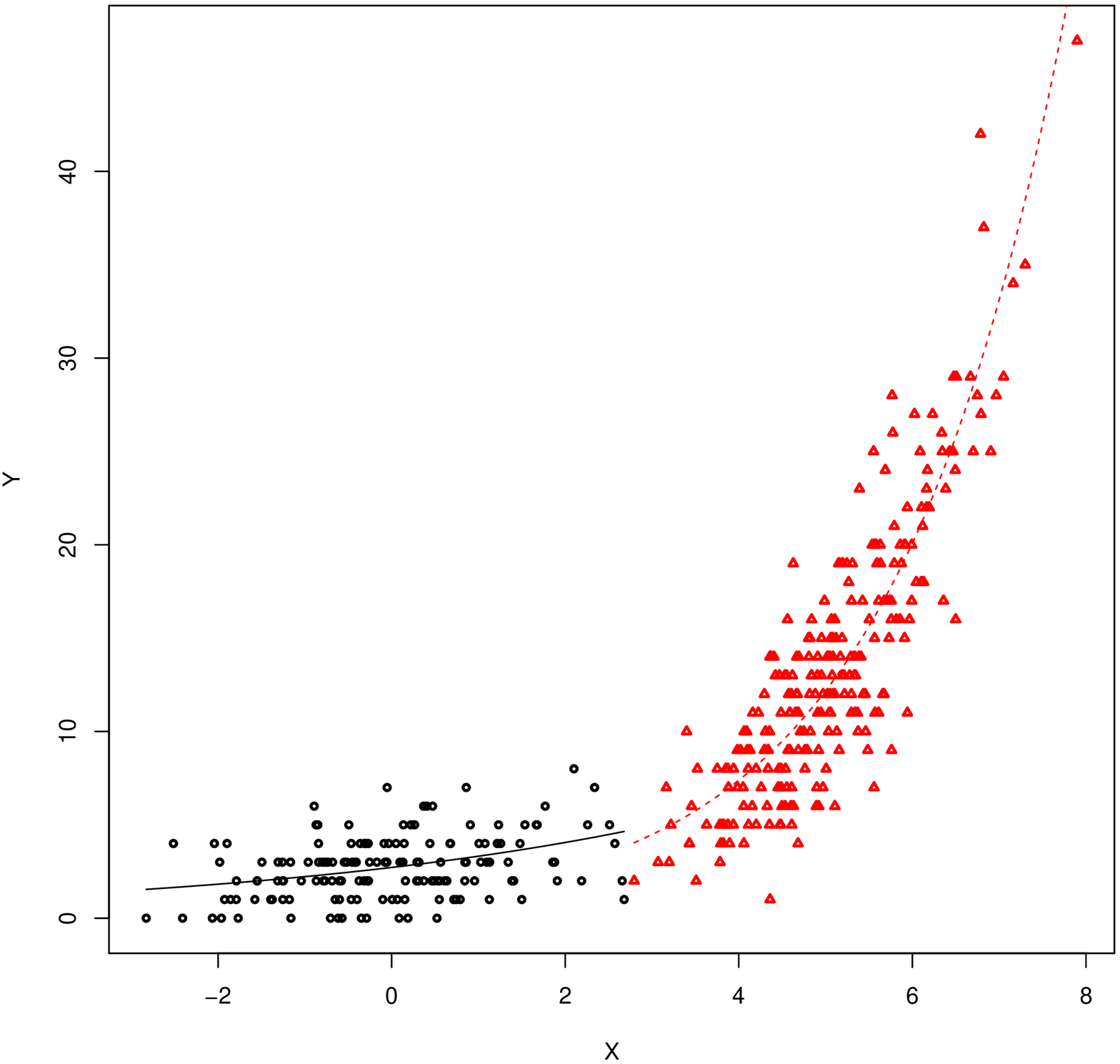}}
\subfigure[Poisson FMR\label{fig:scatterFMR}]
{\includegraphics[width=0.49\textwidth]{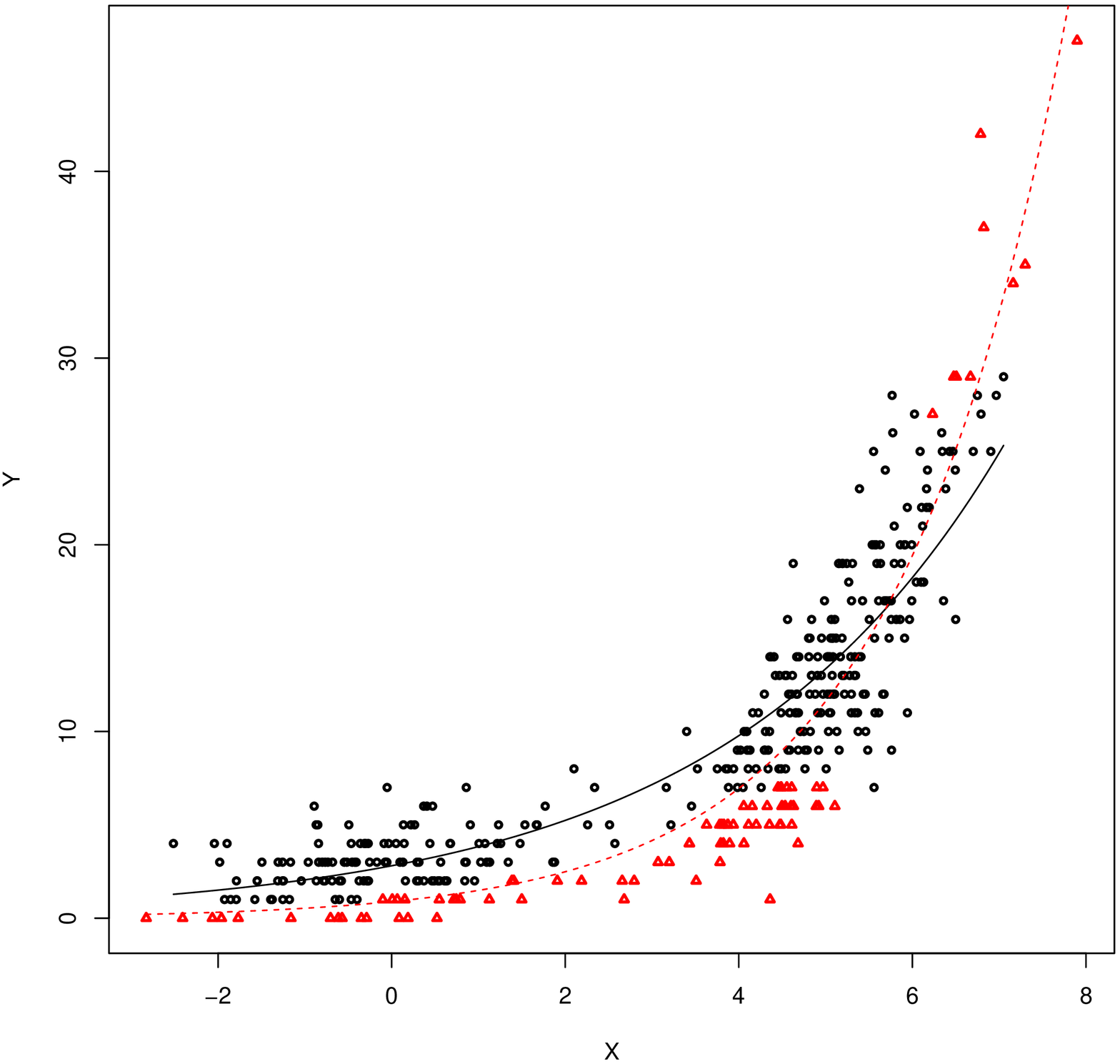}}
\subfigure[Poisson FMRC\label{fig:scatterFMRC}]
{\includegraphics[width=0.49\textwidth]{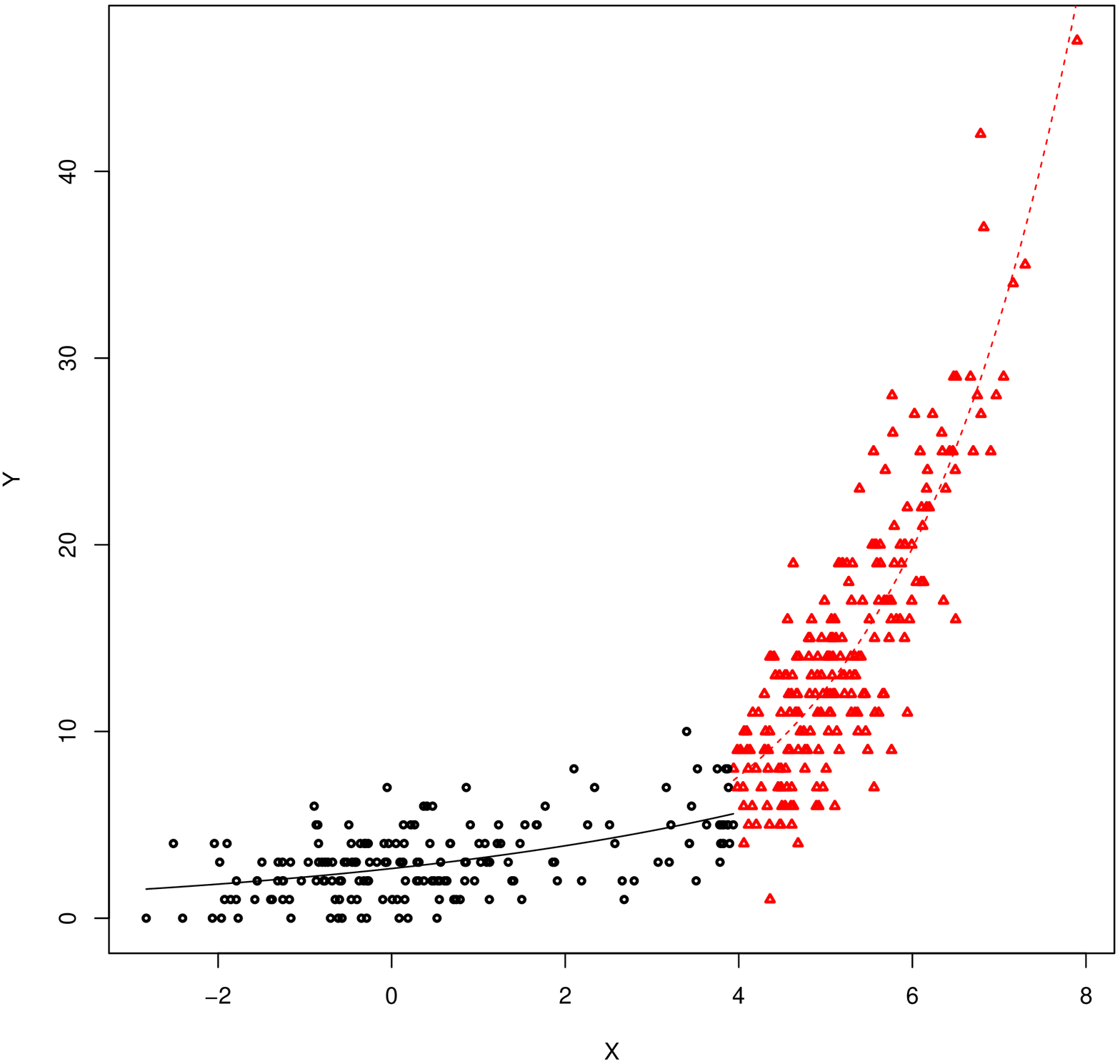}}
\subfigure[Poisson Gaussian CWM\label{fig:scatterCWM}]
{\includegraphics[width=0.49\textwidth]{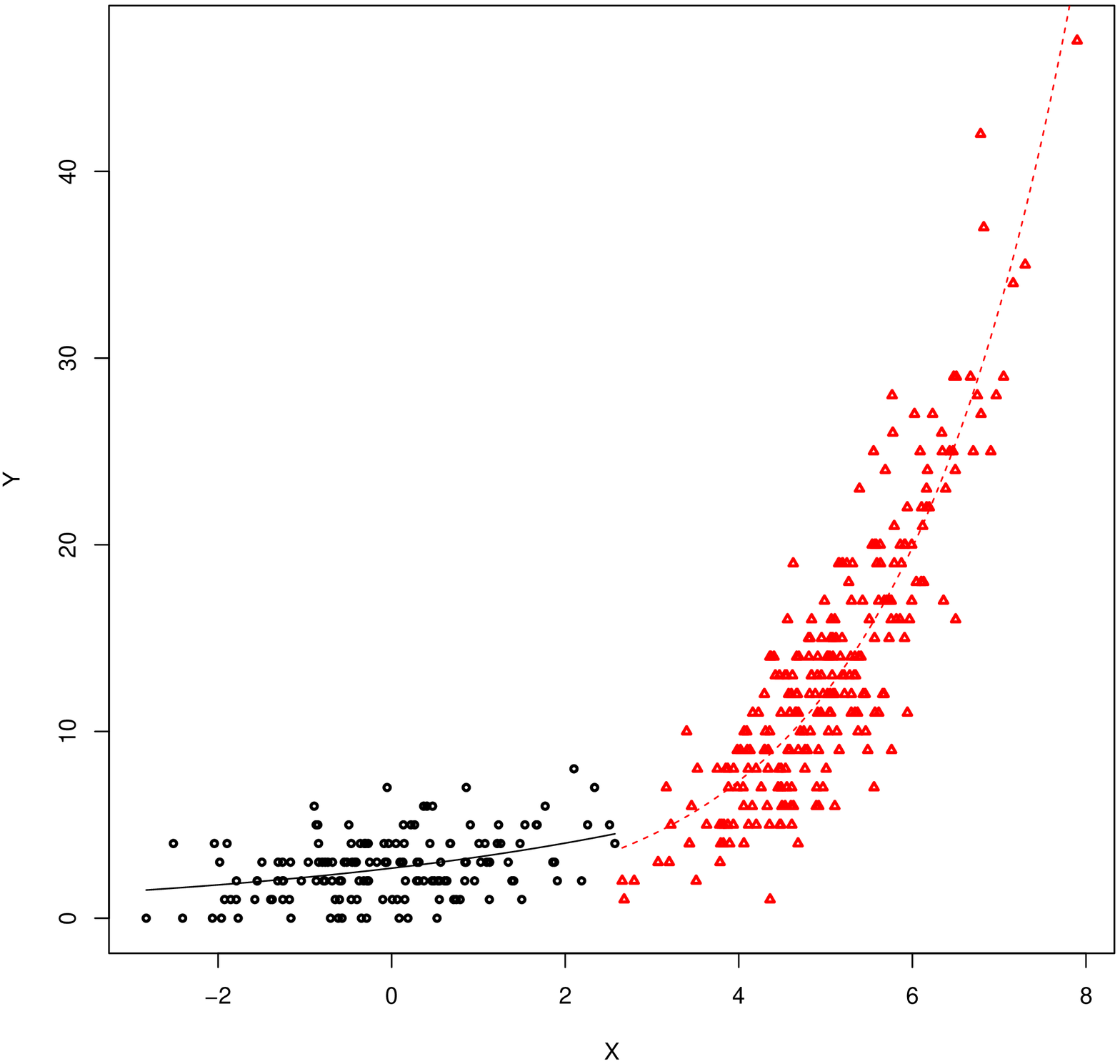}}
\caption{Example \ref{ex:Data from a Poisson GCWM}. 
Scatter plots and Poisson-based models.
}
\label{fig:artificial-CWMscatter}
\end{figure}
\figurename~\ref{fig:PCWM density} displays the joint density from a Poisson CWM.
\begin{figure}[!ht]
\centering
\includegraphics[width=10cm]{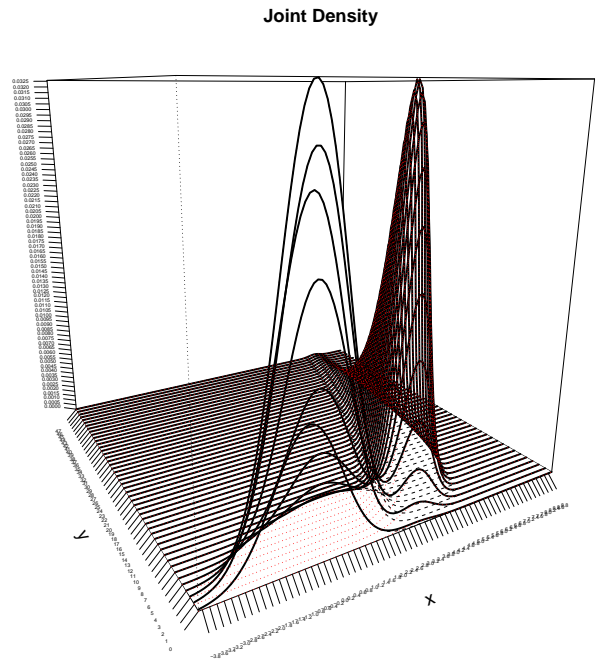} 
\caption{
Example \ref{ex:Data from a Poisson GCWM}. Joint density from the fitted Poisson CWM.}
\label{fig:PCWM density}
\end{figure}
In \tablename~\ref{tab:Clustering performance} we get the ARI and the misclassification error we obtained in the three data modeling.
Here, it is possible to see as the Poisson CWM outperforms the other approaches; in particular,  the Poisson FMR  is not able to capture the underlying group-structure of the data.  
\begin{table}[!ht]
\caption{\it Example \ref{ex:Data from a Poisson GCWM}. Clustering performance of the fitted Poisson-based models}
\label{tab:Clustering performance}
\centering
\begin{tabular}{rrrr}
\hline
	&	FMR	&	FMRC	&	CWM	\\
\hline
misclassError	&	37.50\%	&	6.50\%	&	0.50\%	\\
ARI	&	-0.00378	&	0.75612	&	0.97997	\\
\hline
\end{tabular}
\end{table}
}\end{Ex}

\subsection{Real data}
\label{sec:Real data}

\begin{Ex}[Coupon redemption data.]
\label{ex:coupon redemption data}
{\rm
The following example is based on the coupon redemption data analyzed in \citet{Wede:DeSa:mixtglm:1995}, see also \citet{Wedel:user:2000}. 
The sample size is $N=270$. 
The response variable $Y$ is the number of yogurt coupons redeemed by consumers during a period of 104 weeks and the predictor variable $X$ is the average price paid for ounce. 
We ran both constrained and (unconstrained) Poisson Gaussian CWM, Poisson FMR and Poisson FMRC.  
The number of groups associated with the largest  BIC  is $G=2$. 

Constrained Poisson Gaussian CWM outperforms the unconstrained model (BIC resulted $-1952.12$ and $-1992.64$, respectively) while
Poisson FMR slightly outperforms Poisson FMRC (BIC resulted $-783.71$ and $-788.55$, respectively), where we recall that the BIC values cannot be compared directly  between the two classes of models, due their different nature (see Section~\ref{subsec:discussionCWM->FMR}). 
The constrained Poisson CWM and the Poisson FMR attained very similar values of GCOF and GSD measures of goodness of fit (see \tablename~\ref{tab:GOF_coupon}).
Moreover, they substantially provide the same parameter estimates and the same classification (see \tablename~\ref{tab:parest_coupon} and \figurename~\ref{fig:class_coupon}.  This confirms again theoretical results proved in Section \ref{sec:Parameter estimation}.

\begin{table}[!ht]
\begin{center}
\begin{tabular}{ccrr} 
\hline
group & coefficients & Constrained Poisson Gaussian CWM & Poisson FMR  \\ 
\hline
1  & $\beta_{01}$ & -1.122 & -1.134 \\
  & $\beta_{11}$ & 0.025 & 0.026 \\ 
  & & &  \\
2  & $\beta_{02}$ & -0.435 & -0.445 \\
  & $\beta_{12}$ & 0.250 & 0.250 \\ 
\hline 
\end{tabular}
\end{center}
\caption{
Coupon redemption data.
Coefficients of Constrained Poisson Gaussian CWM and Poisson FMR.}
\label{tab:parest_coupon}
\end{table}

\begin{figure}[!ht] 
	\begin{center}
		\includegraphics[height=8 cm, width=8 cm]{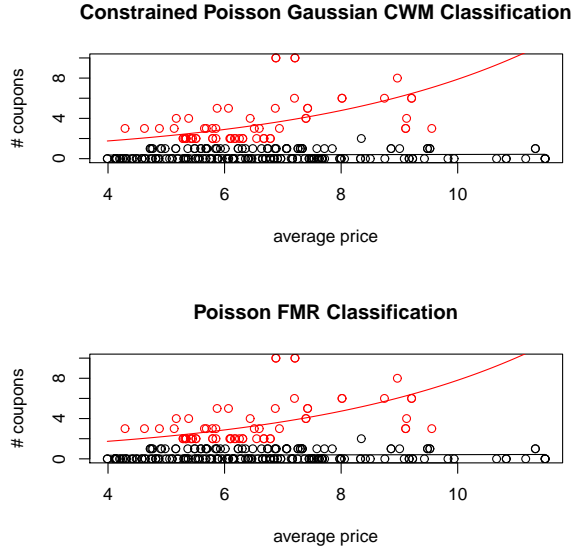}
	\end{center}
	\caption{
	Coupon redemption data.  
	Reclassified data (constrained Poisson Gaussian CWM and Poisson FMR).
	}
	\label{fig:class_coupon}
\end{figure}

\begin{table}
\begin{center}
\begin{tabular}{cccc}
\hline
  & Constrained Poisson Gaussian CWM & Poisson FMR \\ \hline
$GGOF$ & 1.126 & 1.127 \\
$GSD$ & 279.268 & 278.886 \\
\hline
\end{tabular}
\end{center}
\caption{\rm Coupon redemption data. Generalized weighted Pearson chi-square statistic and generalized scaled deviance in Constrained Poisson Gaussian CWM and Poisson FMRC.}\label{tab:GOF_coupon}
\end{table}

}\end{Ex}

\begin{Ex}[Patent data.]
\label{ex:patent data}
{\rm
Patent data have been studied in \citet{WCP:anal:1998} and are available in the R \texttt{Flexmix} library. 
They consist of $N=70$ observations on patent applications and R\&D spending in millions of dollars from pharmaceutical and biomedical companies in 1976.
The number of patent applications is the response variable and the R\&D spending is the continuous predictor variable. 
The number of groups associated with the largest BIC is $G=3$. 

Here, the unconstrained Poisson Gaussian CWM outperforms the constrained model (BIC resulted -774.57 and -793.36, respectively), while Poisson FMRC slightly outperforms Poisson FMR (BIC resulted -438.19 and -441.06, respectively). 
The unconstrained Poisson CWM and the Poisson FMRC attained very similar values of GCOF and GSD measures of goodness of fit (see \tablename~\ref{tab:GOF_patent}).
Again, they substantially provide the same parameter estimates and the same classification (see \tablename~\ref{tab:parest_patent} and \figurename~\ref{fig:class_patent}).

\begin{table}[ht!]
\begin{center}
\begin{tabular}{ccrr} 
\hline
group & coefficients & Poisson Gaussian CWM & Poisson FMRC  \\ 
\hline
1  & $\beta_{01}$ & 2.383 & 2.396 \\
  & $\beta_{11}$ & 0.569 & 0.567 \\ 
  & & &  \\
2  & $\beta_{02}$ & 0.782 & 0.803 \\
  & $\beta_{12}$ & 0.825 & 0.821 \\ 
	& & &   \\
3  & $\beta_{03}$ & -1.812 & -1.801 \\
  & $\beta_{13}$ & 1.410 & 1.404   \\ 
\hline 
\\
\end{tabular}
\end{center}
\caption{
Patent data. 
Coefficients estimates of Poisson Gaussian CWM and Poisson FMRC.}
\label{tab:parest_patent}
\end{table}

\begin{figure}[ht!] 
	\begin{center}
		\includegraphics[height=8 cm, width=8 cm]{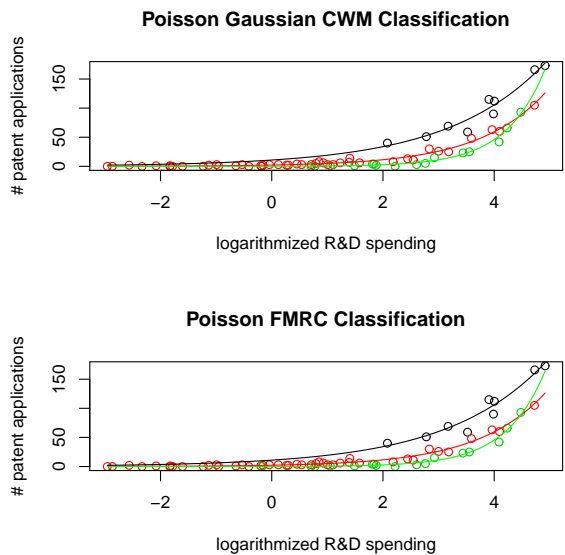}
	\end{center}
	\caption{
	Patent data. 
	Reclassified data (Poisson Gaussian CWM and Poisson FMRC).
	}
	\label{fig:class_patent}
\end{figure}


\begin{table}[!ht]
\begin{center}
\begin{tabular}{cccc}
\hline
  & Poisson Gaussian CWM & Poisson FMRC  \\ 
\hline
GGOF & 1.283 & 1.277   \\
GSD & 89.659 & 87.339   \\
\hline
\end{tabular}
\end{center}
\caption{
Patent data. 
Generalized weighted Pearson chi-square statistic and generalized scaled deviance in Poisson Gaussian CWM and Poisson FMRC.
}
\label{tab:GOF_patent}
\end{table}
}\end{Ex}



\begin{Ex}[Healthcare data.]
\label{ex:Healtcare data}
{\rm
This case study is based on administrative data concerning the healthcare system in the Italian Lombardy region. 
The sample consists of $N=332$ patients on which ``length of stay'' (in days, count response variable $Y$) and ``age'' (covariate $X$) are measured.
Only living persons at the end of the hospitalization have been considered.
Two Diagnosis-Related Groups (DRGs), corresponding to  the codes 348 and 396 are considered, they  define two groups of size $N_1=164$ and $N_1=168$, respectively.
Figure \ref{fig:SanitaTRUE} displays the scatterplot of the data; moreover, in Figures \ref{fig:Sanita-CWMscatter} (b)-(d) 
the fitted Poisson-based models FMR, FMRC and CWM and the corresponding data clustering are plotted. 
\begin{figure}[!ht]
\centering
\subfigure[TRUE\label{fig:SanitaTRUE}]
{\includegraphics[width=0.49\textwidth]{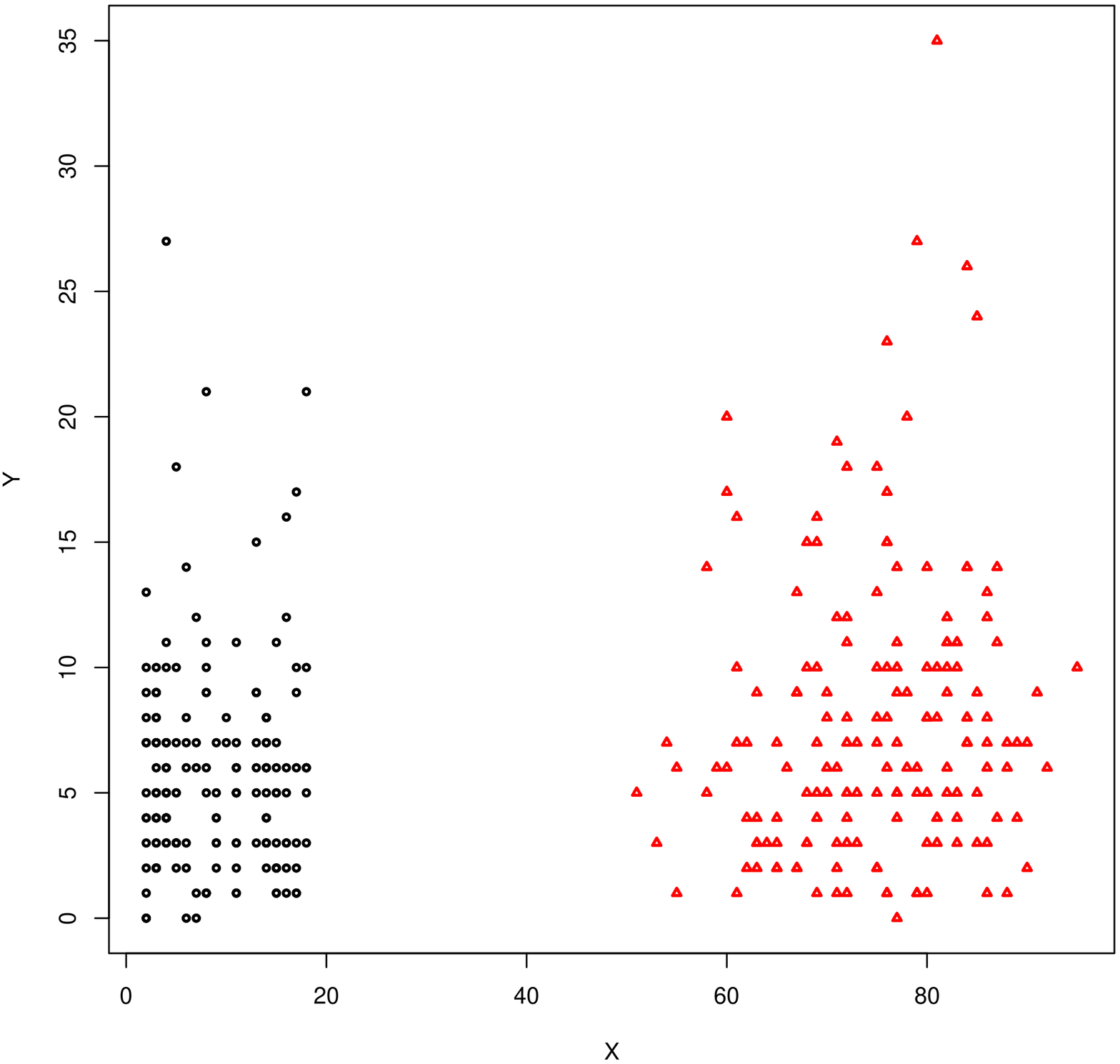}}
\subfigure[Poisson FMR\label{fig:SanitaFMR}]
{\includegraphics[width=0.49\textwidth]{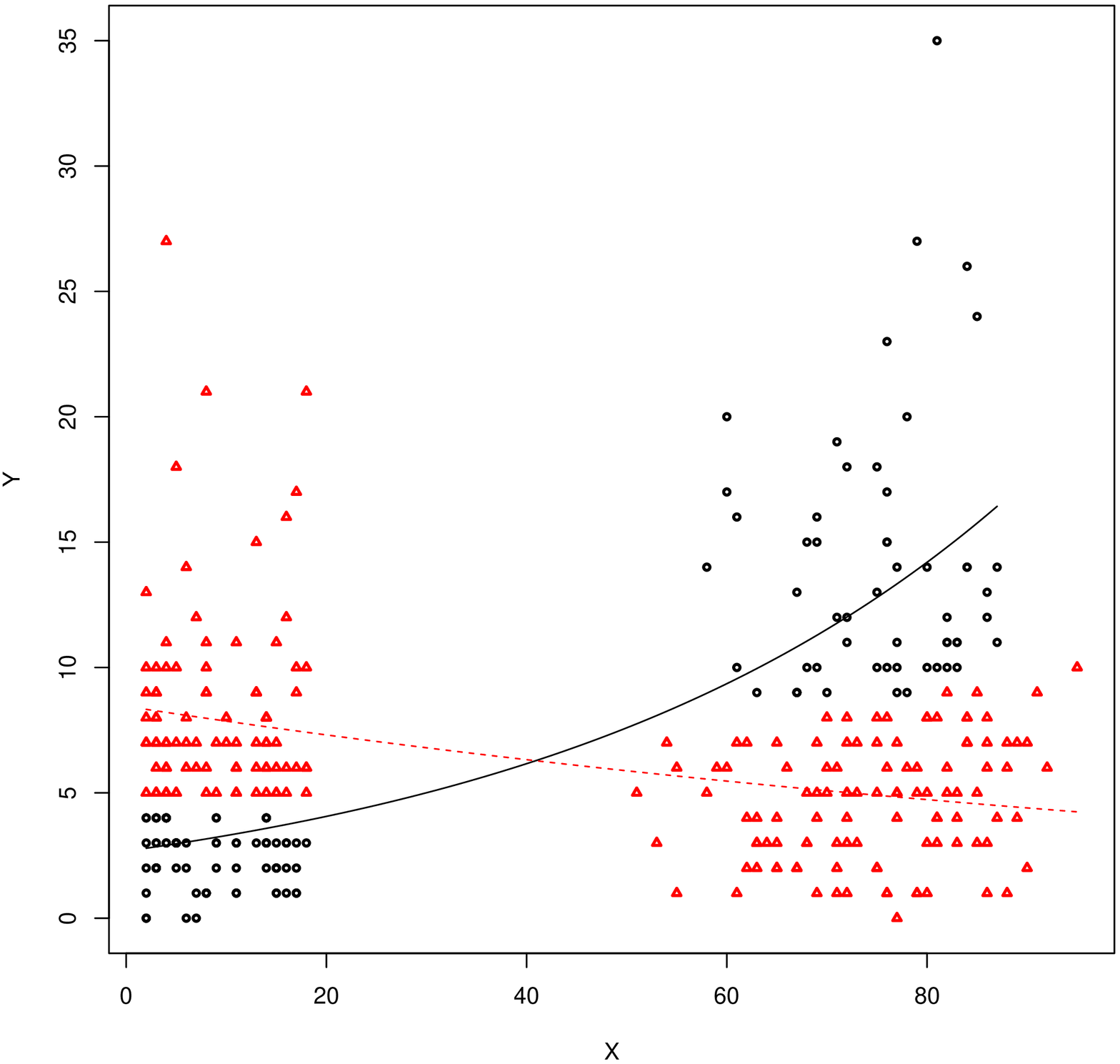}}
\subfigure[Poisson FMRC\label{fig:SanitaFMRC}]
{\includegraphics[width=0.49\textwidth]{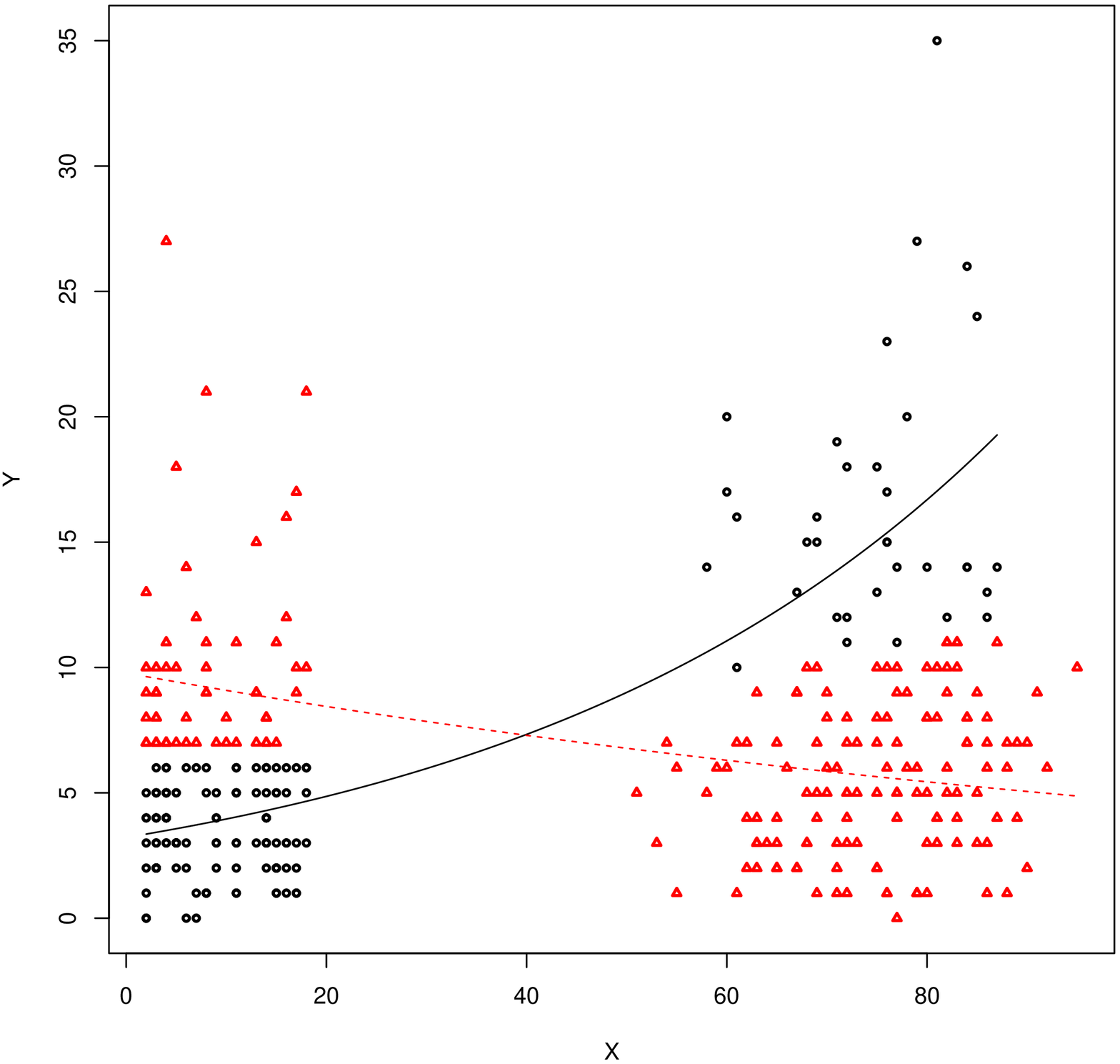}}
\subfigure[Poisson CWM\label{fig:SanitaCWM}]
{\includegraphics[width=0.49\textwidth]{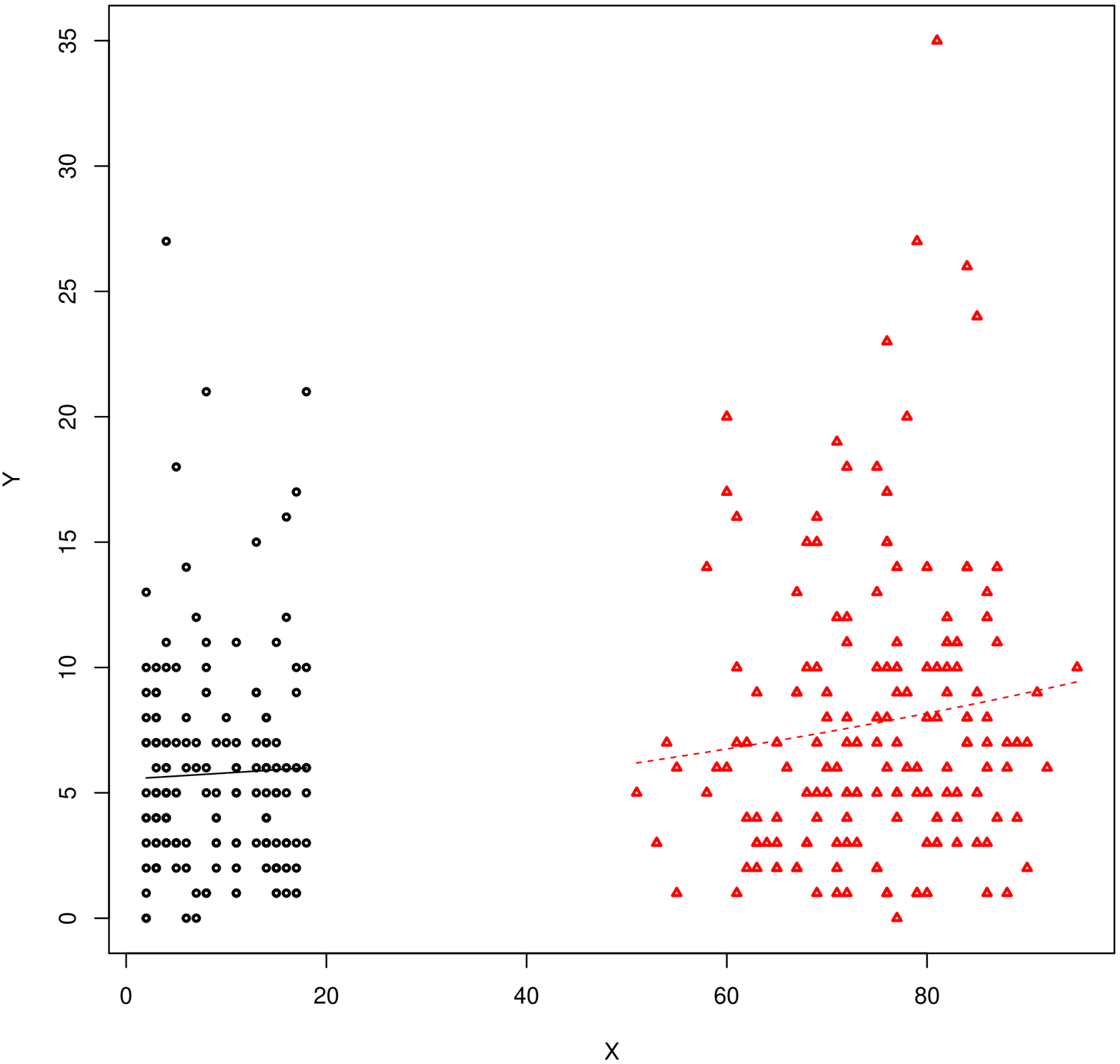}}
\caption{Example \ref{ex:Healtcare data}, DRG: A and B.
Scatter plots and Poisson-based models for the Healthcare data.
}
\label{fig:Sanita-CWMscatter}
\end{figure}
The ARI and the misclassification error  obtained in the three data modeling are given in \tablename~\ref{tab:Sanita Poisson performance}.
The results show that  the Poisson CWM clearly outperforms both FMR and FMRC, and it yields a perfect separation between the two classes.
On the contrary, the Poisson FMR and FMRC are not able to capture the evident underlying group-structure of the data.  
\begin{table}[!ht]
\caption{\it Example \ref{ex:Healtcare data}, DRG: 348 and 396.  
Clustering performance of the fitted Poisson-based models on the Healtcare data}
\label{tab:Sanita Poisson performance}
\centering
\begin{tabular}{rrrr}
\hline
	&	FMR	&	FMRC	&	CWM	\\
\hline
misclassError	&	45.48\%	&	28.01\%	&	0.00\%	\\
ARI	&	0.00531	&	0.19098	&	1.00000	\\
\hline
\end{tabular}
\end{table}
We remark that we carried out other analyses considering patients based on other pairs DRGs. Here, the Poisson 
CWM outperformed both FMR and FMRC, even if in many cases it didn't yield a perfect separation between classes. The results obtained in clustering 
data coming from  other pairs of DRGs  are 
given in Table \ref{tab:Sanita Poisson performance 2}.
\begin{table}[!ht]
\caption{\it Example \ref{ex:Healtcare data}.  
Clustering performance of the fitted Poisson-based models on the Healtcare data (other pairs of DRG).}
\label{tab:Sanita Poisson performance 2}
\centering
\begin{tabular}{ccrrr}
\hline
Couples of DRGs &	&FMR	&	FMRC	&	CWM	\\
\hline
$\left(68,317\right)$	&	&34.67\%	&	37.00\%	&	\textbf{30.00\%}	\\
$\left(345,68\right)$	&	&41.97\%	&	43.00\%	&	\textbf{34.47\%}	\\
$\left(345,168\right)$	&	&43.81\%	&	44.48\%	&	\textbf{36.12\%}	\\
\hline
\end{tabular}
\end{table}

}\end{Ex}

\subsection{Discussion}

\begin{figure}[!ht]
\centering
\subfigure[Example \ref{ex:Binomial-based models}. Binomial-based model.]
{\includegraphics[width=0.49\textwidth]{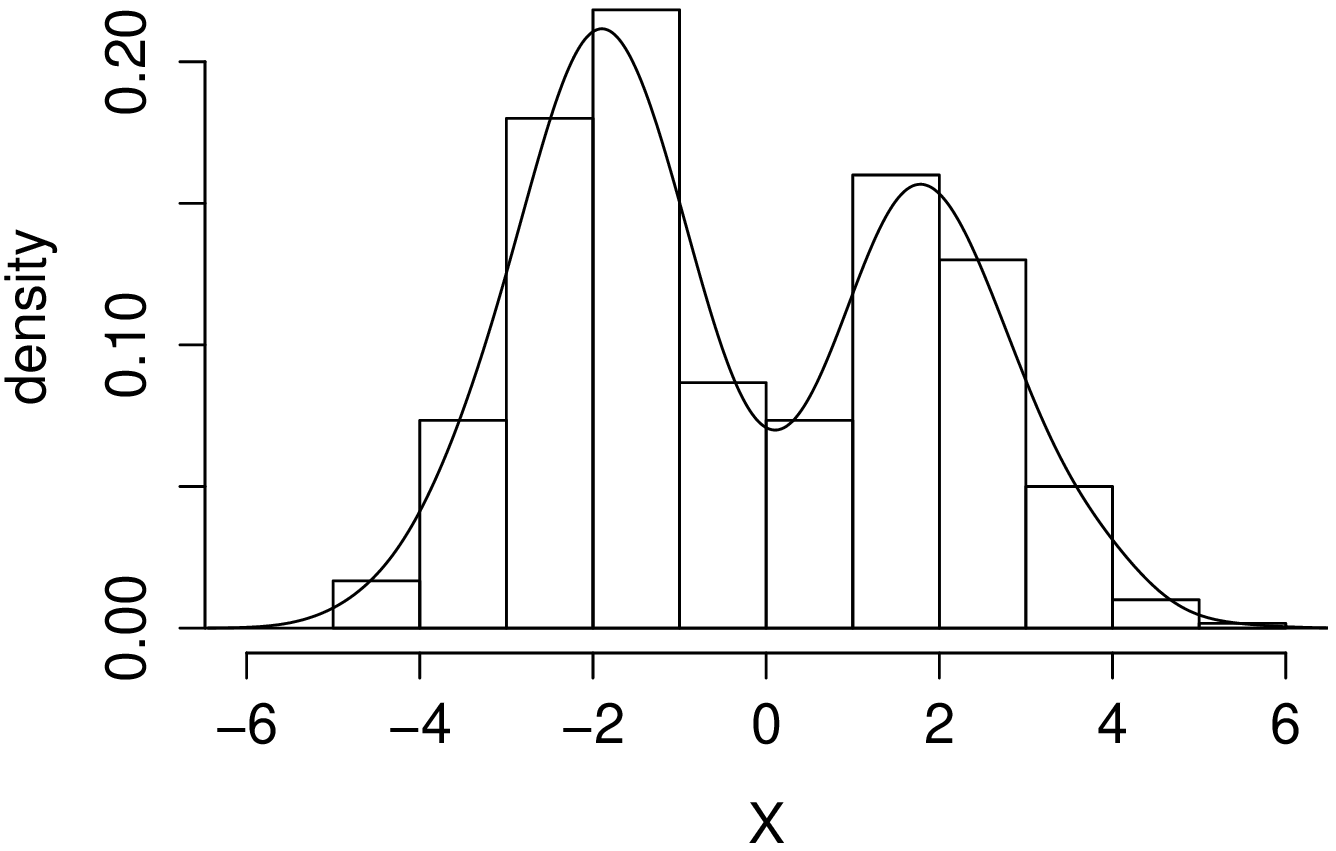}}
\subfigure[Example \ref{ex:Binomial-based models2}. Binomial-based model.]
{\includegraphics[width=0.49\textwidth]{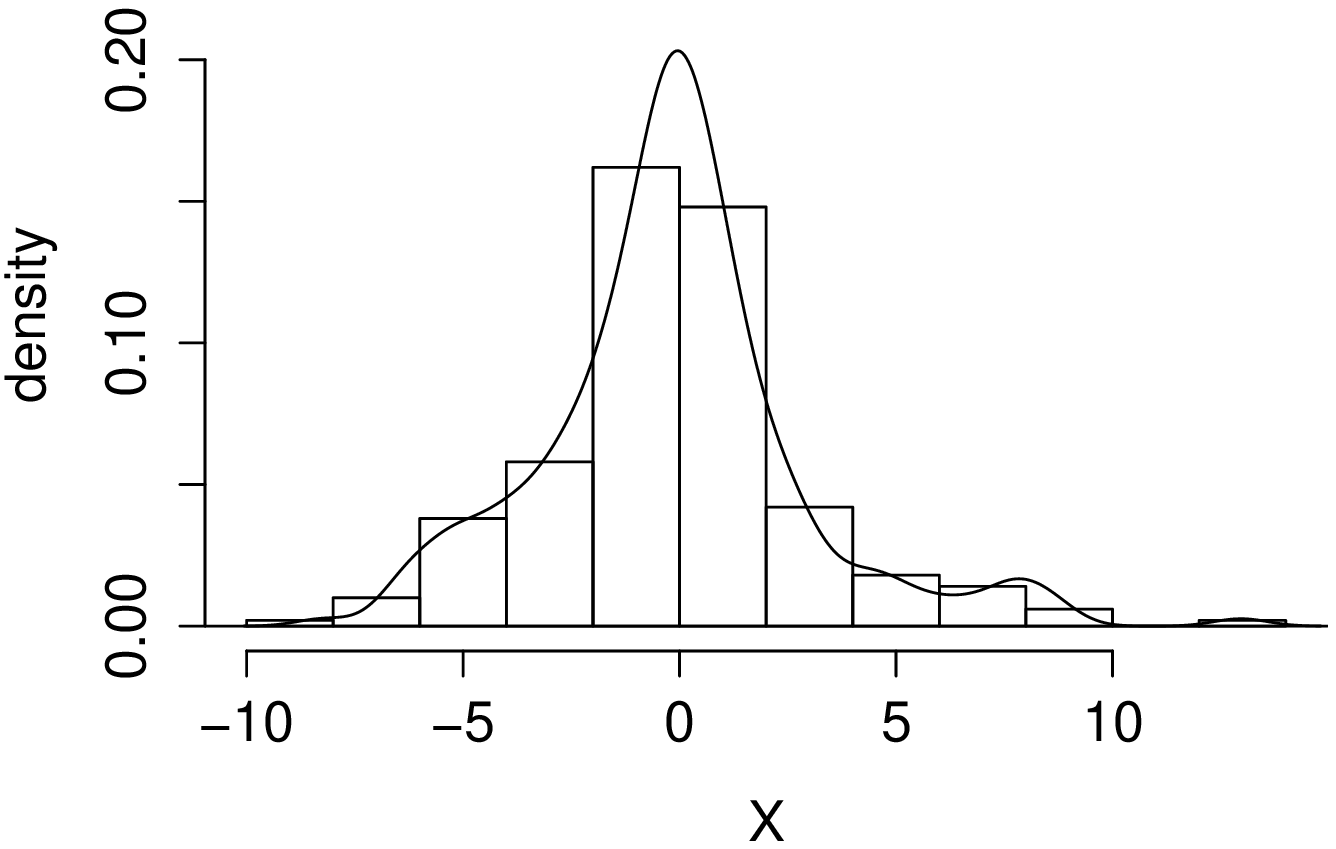}}
\subfigure[Example \ref{ex:Data from a Poisson GCWM}. Poisson-based model.]
{\includegraphics[width=0.49\textwidth]{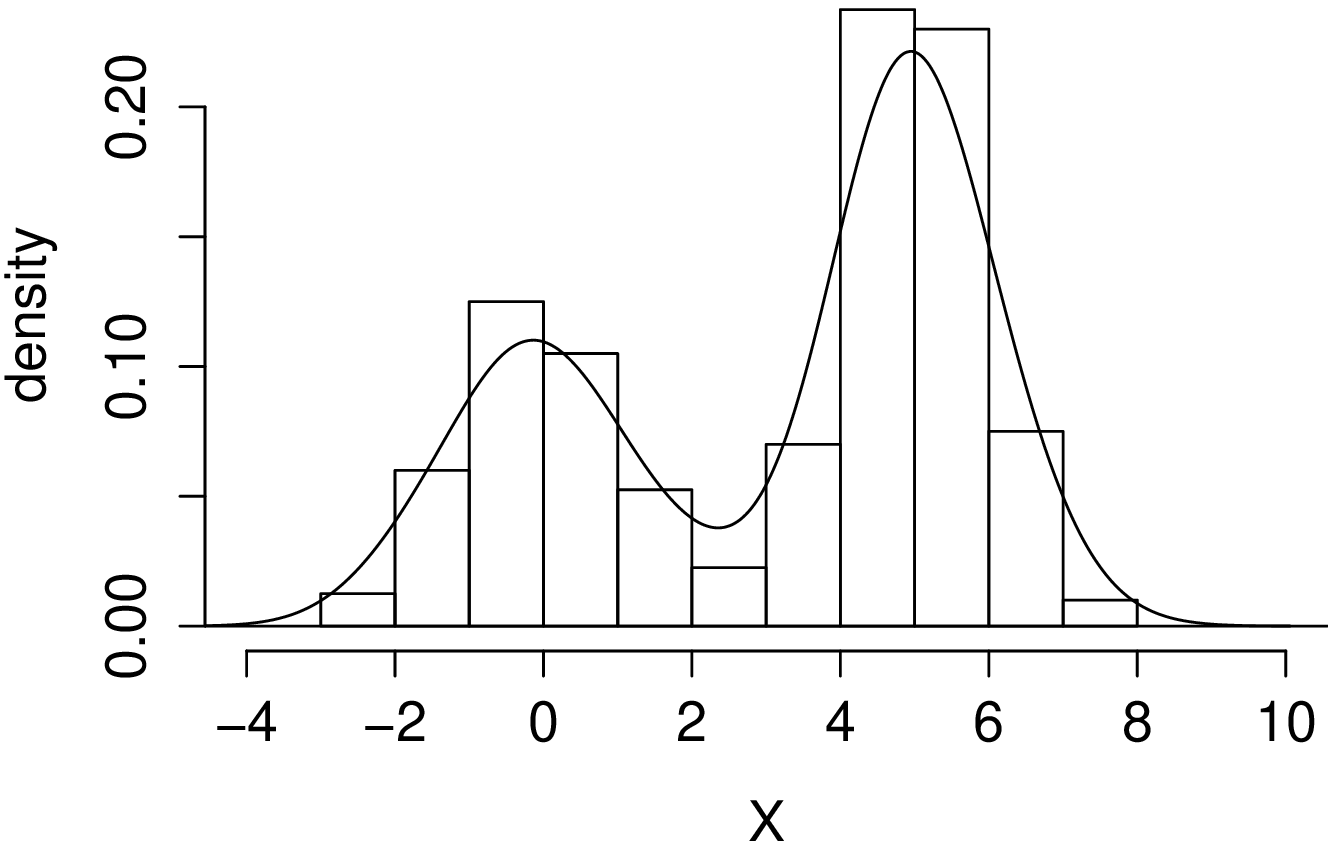}}
\caption{Simulated data examples. Distributions of the $X$ variable.}
\label{fig:simulated X}
\end{figure}

\begin{figure}[!ht]
\centering
\subfigure[Example \ref{ex:coupon redemption data}: Coupon data]
{\includegraphics[width=0.49\textwidth]{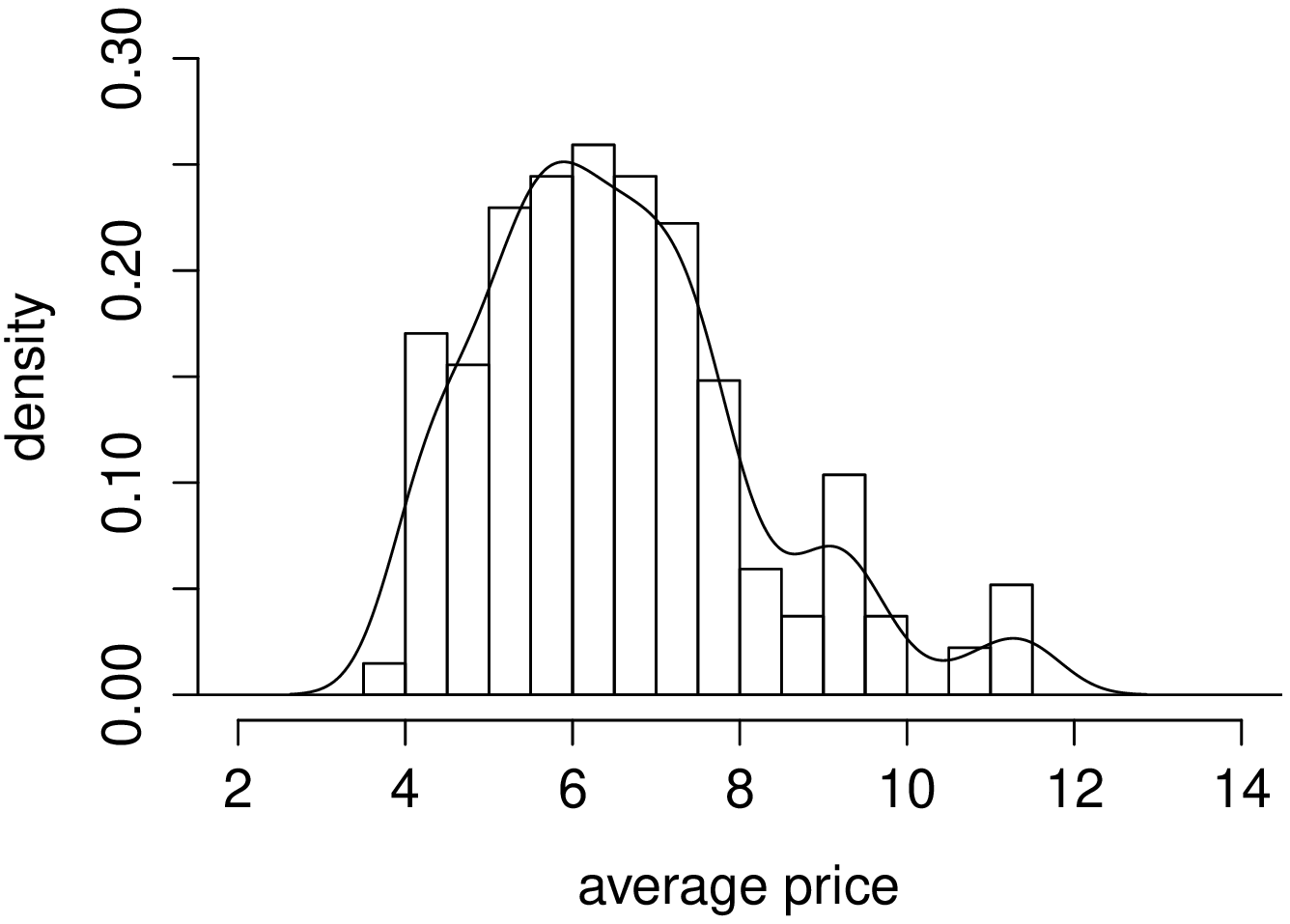}}
\subfigure[Example \ref{ex:patent data}: Patent data]
{\includegraphics[width=0.49\textwidth]{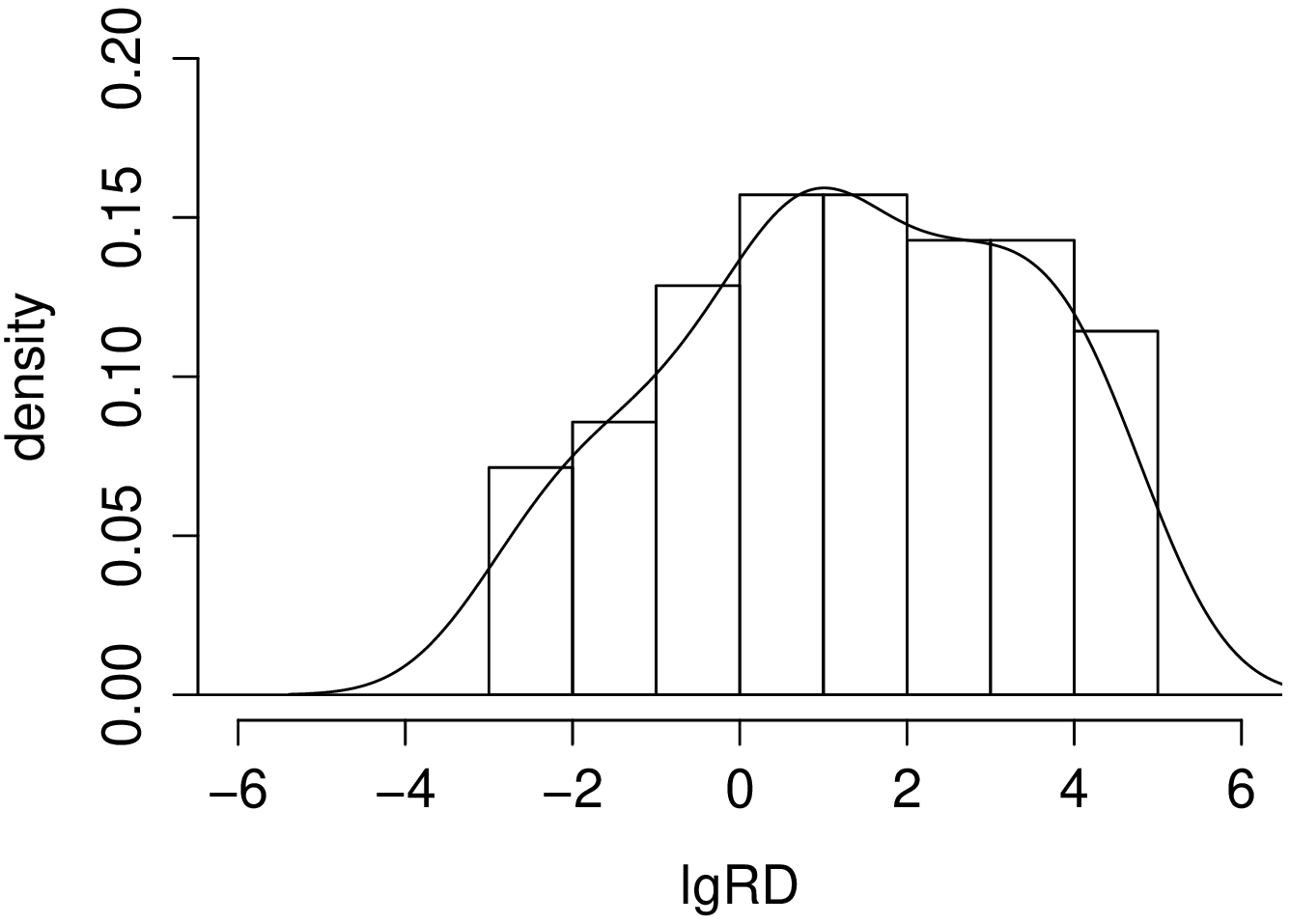}}
\subfigure[Example \ref{ex:Healtcare data}: Health data]
{\includegraphics[width=0.49\textwidth]{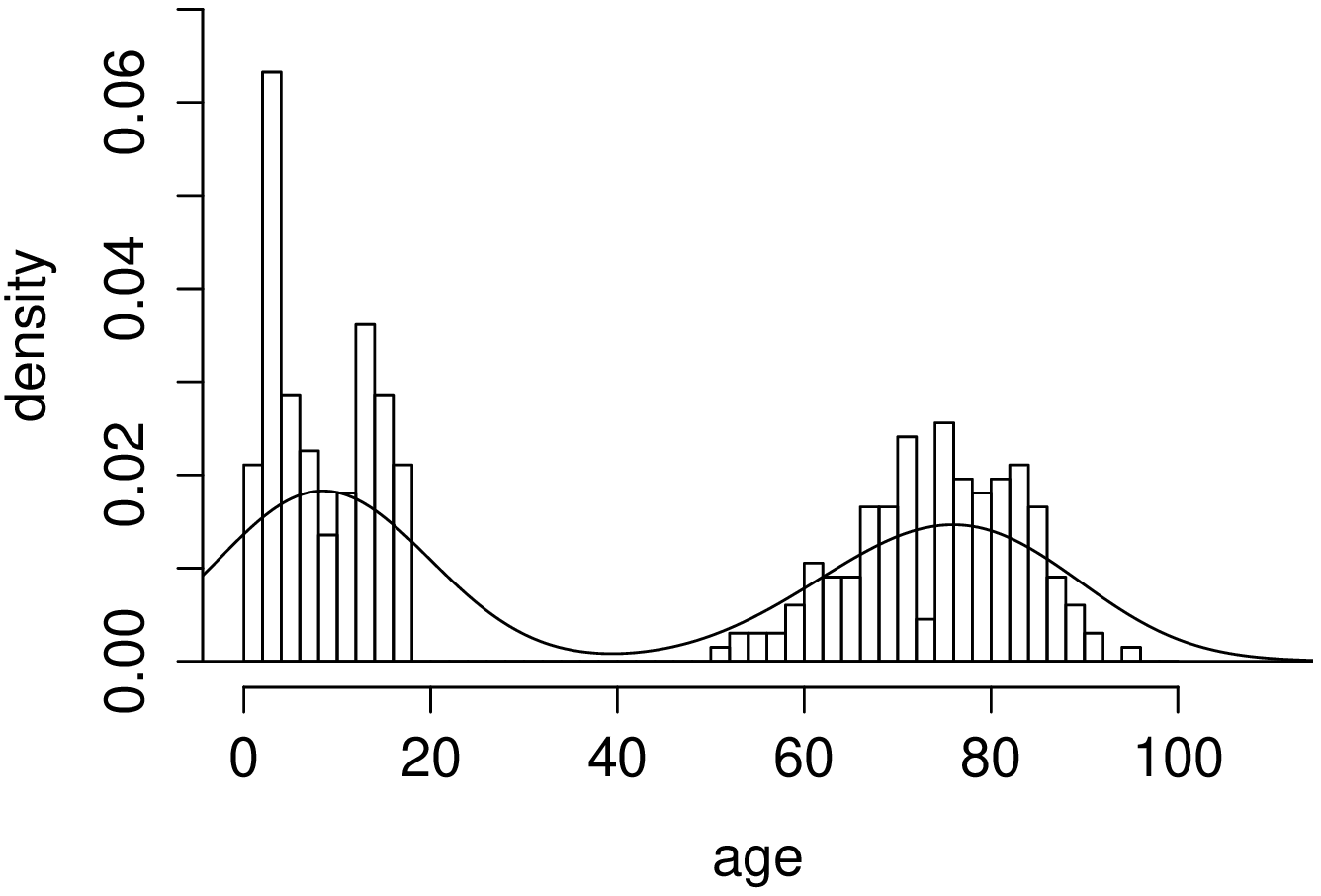}}
\caption{Real data examples. Distributions of the $X$ variable.}
\label{fig:real data X}
\end{figure}

The results of the numerical studies given in this section emphasizes the effectiveness of the GLGCWM in comparison with some  finite mixtures of regression models. Indeed the results of the above examples   can be summarized as follows:
\begin{description}
\item[\underline{Example \ref{Estimates comparison between the MLCM and the constrained GLGCWM}}:] here the theoretical results given in Section~\ref{sec:Parameter estimation}
have been investigated from the numerical point of view and the constrained Poisson GCWM reveals to be a very good approximation for the Poisson GFMR, regardless from the distribution of $X$.
\item[\underline{Example \ref{ex:Binomial-based models}}:] here FMRC strongly outperformed FMR and CWM gave comparable results obtained using FMRC. 
\item[\underline{Example \ref{ex:Binomial-based models2}}:] here FMR and FMRC gave comparable results while CWM strongly outperformed both FMR and FMRC. 
\item[\underline{Example \ref{ex:Data from a Poisson GCWM}}:] here FMRC strongly outperformed FMR and CWM outperfomed FMRC. 
\item[\underline{Example \ref{ex:coupon redemption data}}:] here FMR slightly outperformed FMRC and CWM gave comparable results obtained using FMR. 
\item[\underline{Example \ref{ex:patent data}}:] here FMRC slightly outperformed FMR and CWM gave comparable results obtained using FMRC. 
\item[\underline{Example \ref{ex:Healtcare data}}:] here CWM  outperformed both FMR and FMRC. 
\end{description}
In order to deepen such results, in Figures \ref{fig:simulated X} and \ref{fig:real data X} we show the
distribution along the covariate $X$ for both simulated and real data. Figures \ref{fig:simulated X}(a, c) (Examples \ref{ex:Binomial-based models}
and \ref{ex:Data from a Poisson GCWM}) and Figure \ref{fig:real data X}(c) (Example \ref{ex:Healtcare data}) 
highlight that CWM yields better performance than FMR when   the covariates  present a clear group-structure; 
\figurename~\ref{fig:simulated X}(b) (Examples \ref{ex:Binomial-based models2})  
highlights that CWM yields better performance than FMRC when distributions of the covariates are overlapped with different ranges.
This confirm the previous results  obtained in \citet{Ingr:Mino:Vitt:Loca:2012}.
On the contrary, when there is no a  group-structure in the covariates, then CWM, FMR and FMRC give comparable results, see 
Figures \ref{fig:real data X}(a,b) (Examples \ref{ex:coupon redemption data} and \ref{ex:patent data}).  

\section{Concluding remarks}
\label{sec:Conclusions}
In this paper we have introduced the generalized linear Gaussian CWM (GLGCWM) and we have shown that they are  quite
flexible statistical models.
Such models allow modeling categorical variables depending on numerical covariates based on data  
coming from a heterogeneous population.  Some relationships with finite mixtures of generalized mixture models (with and without concomitants) have also been investigated;
in particular, we have shown that  mixtures of generalized linear models  can be considered as nested models in GLGCWM  even if they have a different structure.  
Applications to real and artificial data have emphasized the effectiveness of the proposal, also in comparison with the other models cited above. 

We remark that in this paper we have considered Gaussian marginal distributions.
However, the
extension to multivariate Student-$t$ is straightforward, and in this case  the parameters of  Student-$t$ local densities
can be estimated according to mixtures of multivariate $t$ distributions, see  e.g Section 7.5 in \citet{McLa:Peel:fini:2000}.
In this direction, we are currently working on further extensions of the models presented here. In this framework, an important
issue concern data modeling by Cluster Weighted approaches when the dimension of the input vector $\bX$ is large. First results, are given in \citet{Sube:Punz:Ingr:McNi:GaussianCWMFactor:2012}.

An  issue which deserves attention for future research  concerns the behaviour of the EM algorithm. It is well known that 
the EM algorithm suffers for local maxima and singularities and  our first simulation studies confirmed previous results  given  in literature, see e.g. \citet{Faria:Soro:2010},  \citet{Ingr:Mino:Vitt:Loca:2012}, \cite{Ingr:Mino:Punz:Mode:2012}, showing that  the performance of the algorithm strongly depend 
on the choice of initial values.  Moreover, suitable constraints on the eigenvalues of the covariance
matrices of the marginal distributions can be imposed in order to run the algorithm in a parameter space with
no singularities and a reduced  number of local maxima, see e.g. \citet{Ingrassia:Rocci:07} for details.

Finally, we point out  that the above results open the perspective of model based clustering of mixed type data coming from distributions with density $p(\bx,y)$, where $\bX$ is a mixed  continuous variable and $Y$ is either continuous or discrete.

\section*{Appendix: Proofs}

\paragraph{Proof of Proposition \ref{prop:CWM vs finite mixture of GLMs}}  
If $\bmu_g, \bSigma_g=\bmu, \bSigma$, $g=1,\ldots,G$, then \eqref{eq:Generalized Linear Gaussian CWM} becomes
\begin{align*}
p(\bx,y; \btheta) & =  \sum^{G}_{g=1} q(y|\bx; \bbeta_g, \lambda_g) \phi_d(\bx;\bmu, \bSigma) \pi_g 
 =  \phi_d(\bx;\bmu, \bSigma)\sum^{G}_{g=1} q(y|\bx; \bbeta_g, \lambda_g) \pi_g \\ 
& =  \phi_d(\bx;\bmu, \bSigma)f(y|\bx;\bkappa).
\end{align*}
\qed

\paragraph{Proof of Corollary \ref{cor:posteriors CWM vs posteriors finite mixture of GLMs}}  
If $(\bmu_g, \bSigma_g)=(\bmu, \bSigma)$, $g=1,\ldots,G$, then the posterior probability that the generic observation $(\bx',y)'$ belongs to $\Omega_g$ from model \eqref{eq:Generalized Linear Gaussian CWM} specifies as
\begin{align*}
p(\Omega_g|\bx,y) & =  \frac{q(y|\bx; \bbeta_g, \lambda_g) \phi_d(\bx;\bmu, \bSigma)\pi_g}{\displaystyle\sum^{G}_{j=1} q(y|\bx;\bbeta_j, \lambda_j) \phi_d(\bx;\bmu, \bSigma)\pi_j} =  \frac{q(y|\bx;\bbeta_g, \lambda_g) \phi_d(\bx;\bmu, \bSigma)\pi_g}{\phi_d(\bx;\bmu, \bSigma) \displaystyle\sum^{G}_{j=1} q(y|\bx; \bbeta_j, \lambda_j) \pi_j} \\
& =  \frac{q(y|\bx; \bbeta_g, \lambda_g)\pi_g}{\displaystyle\sum^{G}_{j=1} q(y|\bx;\bbeta_j, \lambda_j) \pi_j}, \quad g=1,\ldots,G,
\end{align*}
which coincides with \eqref{eq:posteriors from a finite mixture of GLMs}.
\qed

\paragraph{Proof of Proposition \ref{prop:CWMLG-FMLRC}}  
From \eqref{eq:Generalized Linear Gaussian CWM} we get:
\begin{align*}
p(\bx,y;\btheta) 
& =  \sum^{G}_{g=1} q(y|\bx; \bbeta_g, \lambda_g) \phi_d(\bx; \bmu_g, \bSigma)\pi   
 =  p(\bx;\bpsi)\sum^{G}_{g=1} p(y|\bx;\bbeta_g, \lambda_g)\frac{\phi_d(\bx;\bmu_g,\bSigma) \pi}{p(\bx;\bpsi)} \\
& =  p(\bx;\bpsi)\sum^{G}_{g=1} q(y|\bx;\bbeta_g, \lambda_g) \frac{\displaystyle\exp\left[-\frac{1}{2}(\bx-\bmu_g)'\bSigma^{-1}(\bx-\bmu_g)\right] }{\displaystyle\sum_{j=1}^{G}\exp\left[-\frac{1}{2}(\bx-\bmu_j)'\bSigma^{-1}(\bx-\bmu_j)\right] },
\end{align*}
where
\begin{align}
& \frac{\displaystyle\exp\left[-\frac{1}{2}(\bx-\bmu_g)'\bSigma^{-1}(\bx-\bmu_g)\right] }{\displaystyle\sum_{j=1}^{G}\exp\left[-\frac{1}{2}(\bx-\bmu_j)'\bSigma^{-1}(\bx-\bmu_j)\right] }  \notag \\
& \qquad = \frac{1}{1+\displaystyle\sum_{j\neq g}\exp\left[-\frac{1}{2}(\bx-\bmu_j)'\bSigma^{-1}(\bx-\bmu_j)+\frac{1}{2}(\bx-\bmu_g)'\bSigma^{-1}(\bx-\bmu_g)\right]}  \notag \\
& \qquad = \frac{1}{1+\displaystyle\sum_{j\neq g}
\exp\left[(\bmu_j-\bmu_g)' \bSigma^{-1}\bx -\frac{1}{2}(\bmu_j+\bmu_g)'\bSigma^{-1}(\bmu_j -\bmu_g)\right]}
\notag \\
& \qquad = \frac{1}{1+\displaystyle\sum_{j\neq g}
\exp\left[\bmu_j'\bSigma^{-1}\bx-\frac{1}{2}\bmu_j'\bSigma^{-1}\bmu_j-(\bmu_g'\bSigma^{-1}\bx-\frac{1}{2}\bmu_g'\bSigma^{-1}\bmu_g)
\right]}
\notag \\
& \qquad = \frac{\displaystyle\exp(-\frac{1}{2}\bmu_g'\bSigma^{-1}\bmu_g+\bmu_g'\bSigma^{-1}\bx) }{\displaystyle\sum_{j=1}^{G}\exp(-\frac{1}{2}\bmu_j'\bSigma^{-1}\bmu_j+\bmu_j'\bSigma^{-1}\bx) }.
\label{expGauss}
\end{align}
If we set
\begin{equation}
\alpha_{g0}=-\frac{1}{2}\bmu_g'\bSigma^{-1}\bmu_g \quad \text{and} \quad \balpha'_{g1}=\bmu_g'\bSigma^{-1},\quad g=1,\ldots,G,
\label{eq:multinomial parameters}	
\end{equation}
we recognize that \eqref{expGauss} can be written in form \eqref{mult_log}.
This completes the proof. \qed

\paragraph{Proof of Corollary \ref{cor:CWMLG-FMLRC}}  
If $\bX|\Omega_g \sim N_d(\bmu_g,\bSigma)$ and $\pi_g=\pi=1/G$, $g=1,\ldots,G$, then the posterior probability that the generic observation $(\bx',y)'$ belongs to $\Omega_g$ from model \eqref{eq:Generalized Linear Gaussian CWM} specifies as
\begin{align*}
p(\Omega_g|\bx,y) &=\frac{q(y|\bx; \bbeta_g, \lambda_g) \phi_d(\bx;\bmu_g,\bSigma) }{\displaystyle\sum^{G}_{j=1}q(y|\bx; \bbeta_j, \lambda_j) \phi_d(\bx;\bmu_j,\bSigma) }\\
& = \frac{q(y|\bx; \bbeta_g, \lambda_g) \exp\left[-\frac{1}{2}(\bx-\bmu_g)'\bSigma^{-1}(\bx-\bmu_g)\right]}{\displaystyle\sum^{G}_{j=1}q(y|\bx; \bbeta_j, \lambda_j) \exp\left[-\frac{1}{2}(\bx-\bmu_j)'\bSigma^{-1}(\bx-\bmu_j)\right] }
\end{align*}
which can be written  as
\begin{equation}
p(\Omega_g|\bx,y) = \frac{q(y|\bx; \bbeta_g, \lambda_g)\exp(\alpha_{g0}+\balpha'_{g1}\bx)}{\displaystyle\sum_{j=1}^G  q(y|\bx;\bbeta_j, \lambda_j)  \exp(\alpha_{j0}+\balpha_{j1}'\bx)},
\label{postFMLRCb}
\end{equation}
where $\alpha_{g0}$ and $\balpha_{g1}$ are specified as in \eqref{eq:multinomial parameters}. \qed

\paragraph{Proof of Proposition \ref{CWM->FMR}}  
In order to prove the proposition, it is sufficient to show that, under the assumption that $\bmu_g, \bSigma_g=\bmu, \bSigma$, the terms $\cL_{1c} (\bxi)$ and $\cL_{3c} (\bpi)$ in  \eqref{llikCWM} do
not depend on $\bmu, \bSigma$. Indeed, if $ \bmu_g, \bSigma_g = \bmu, \bSigma $ then the complete-data log-likelihood function becomes:
\begin{align}
\cL_c (\btheta; \sbX, \by)  
& = \sum_{n=1}^N \sum_{g=1}^G  \left[ z_{ng} \ln q(y_n|\bx_n; \bbeta_g, \lambda_g) 
+ z_{ng} \ln  \phi_d(\bx_n;\bmu, \bSigma ) + z_{ng} \ln  \pi_g \right] \notag \\
& = \cL_{1c} (\bxi) + \cL_{2c} (\bpsi^*) + \cL_{3c} (\bpi), \label{llikcCWMg=}
\end{align}
where $\bpsi^*=(\bmu, \bSigma)$ and  $\cL_{2c} (\bpsi)$ in \eqref{llikCWM} is now replaced by 
\begin{align*}
\cL_{2c} (\bpsi^*) 
& = \sum_{n=1}^N \ln  \phi_d(\bx_n;\bmu, \bSigma ) ,
\end{align*}
since $\sum_{g=1}^G  z_{ng}=1$ for $n=1, \ldots, N$.  
Moreover, since $(\bmu_g, \bSigma_g)=(\bmu, \bSigma)$ for $g=1, \ldots, G$, then the posterior probability  \eqref{eq:CWM posteriors} reduces to
\begin{equation*}
 p(\Omega_g|\bx,y) =\frac{q(y|\bx; \bbeta_g, \lambda_g) \pi_g}{\sum^{G}_{j=1}q(y|\bx; \bbeta_j, \lambda_j) \pi_j}.
\end{equation*}
Thus, $z_{ng}$ does not depend on $\phi_d(\bx_n;\bmu_g, \bSigma_g )$ and neither does the  term $\cL_{3c} (\bpi)$.  
In summary, the maximization of \eqref{llikcCWMg=} can be attained by independently maximizing the three terms $\cL_{1c} (\bxi),  \cL_{2c} (\bpsi^*)$ and $\cL_{3c} (\bpi)$ and hence, 
the maximization of (\ref{llikcFMR}) and (\ref{llikcCWMg=})  leads to the same estimates of $(\bxi, \bpi)$. This completes the proof. \qed

\paragraph{Proof of Proposition \ref{CWM->FMRC}}  
In order to prove the result, it is sufficient to show that if  $\bSigma_g=\bSigma$ and $\pi_g=1/G$, $g=1, \ldots, G$, 
then the terms $\cL_{1c} (\bxi)$ and $\cL_{3c} (\bpi)$ in \eqref{llikCWM} do not depend on $(\bmu_g, \bSigma)$, $g=1, \ldots, G$. Indeed, we have:
\begin{align}
L_c(\btheta; \sbX, \by) & = \prod_{n=1}^N \prod_{g=1}^G q(y_n|\bx_n; \bbeta_g, \lambda_g)^{z_{ng}}  \phi_d(\bx_n; \bmu_g, \bSigma)^{z_{ng}} \pi^{z_{ng}}  \label{likFMRC}  
\end{align}
and taking the logarithm of \eqref{likFMRC}, after some algebra we get
\begin{align}
\cL_c (\btheta; \sbX, \by)  & = \ln L_c(\btheta; \sbX, \by)   \notag \\
& = \sum_{n=1}^N \sum_{g=1}^G  \left[ z_{ng} \ln  q(y_n|\bx_n; \bbeta_g, \lambda_g) + z_{ng} \ln  \phi_d(\bx_n; \bmu_g, \bSigma)  \right] +\pi \notag\\
& = \cL_{1c} (\bxi) + \cL_{2c} (\bpsi^{**}) +\pi,  \label{llikcCWMgSp=}
\end{align}
where $\bpsi^{**}=\{ \bmu_g, \bSigma \,; g=1, \ldots, G\}$ and $\cL_{2c} (\bpsi)$ in \eqref{llikCWM} is now replaced by 
\begin{align*}
\cL_{2c} (\bpsi^{**}) & = \frac{1}{2} \sum_{n=1}^N \sum_{g=1}^G  z_{ng}\left[- p \ln 2\pi - \ln |\bSigma| - (\bx_n-\bmu_g)' \bSigma^{-1} (\bx_n-\bmu_g)\right].
\end{align*}
Once the estimates of $(\bmu_g, \bSigma)$ have been obtained, quantity $p(\Omega_g|\bx, \bxi)$  in \eqref{llikcFMRC} can be obtained immediately
like in \eqref{postFMLRCb}.
This completes the proof. \qed

\bibliographystyle{natbib}
\bibliography{References}

\end{document}